\documentclass[aps,prb,superscriptaddress,twocolumn,showpacs]{revtex4-2}
\usepackage{silence}
\usepackage{amsmath, amssymb, graphicx, array, dcolumn, subfigure, url, color, multirow, comment}

\usepackage[colorlinks=true, linkcolor=blue, citecolor=blue, urlcolor=blue]{hyperref}
\usepackage{amssymb}
\usepackage{epsfig}
\DeclareMathAlphabet{\pazocal}{OMS}{zplm}{m}{n}  
\usepackage{graphicx}
\usepackage{color}
\usepackage{bm}
\usepackage[normalem]{ulem}     
\usepackage{soul}
\usepackage{amsmath}
\usepackage{braket} 
\usepackage{rotating}
\usepackage{multirow} 
\DeclareMathAlphabet{\pazocal}{OMS}{zplm}{m}{n}            
\usepackage{graphicx}
\usepackage{hyperref}
\usepackage{mhchem}
\usepackage{xcolor}

\begin{document}
 
\title{Orbital and Spin Nernst Effects in Monolayers of Transition Metal Dichalcogenides
}

\author{Saikat Saha}
\affiliation{Department of Physics, Indian Institute of Technology Bombay, Mumbai 400076, India}

\author{Arnab Bose}
\affiliation{Department of Electrical Engineering, Indian Institute of Technology Kanpur, Kanpur 208016, India}

\author{Sayantika Bhowal}
\email{sbhowal@iitb.ac.in}
\affiliation{Department of Physics, Indian Institute of Technology Bombay, Mumbai 400076, India}

\begin{abstract}
In recent years, orbitronic effects have attracted growing attention as complementary counterparts to the well-established spintronic phenomena. In this work, we demonstrate that monolayers of transition metal dichalcogenides provide an excellent platform for the observation of the orbital Nernst effect, a relatively less explored phenomenon describing the generation of a transverse orbital current in response to an applied temperature gradient. We show that, similar to its electrical counterpart, viz., the orbital Hall effect, the orbital Nernst effect does not require the presence of spin–orbit coupling. Analytical results based on a low-energy valley model offer key insights into the underlying mechanisms, highlighting in particular the crucial role of electronic states at the Fermi energy for the emergence of this effect. The inclusion of spin–orbit coupling further gives rise to a spin Nernst effect, which scales with the strength of spin–orbit coupling and vanishes in its absence. We substantiate our analytical findings with full Brillouin-zone tight-binding results for two representative systems, monolayer 2H-MoS$_2$ and 2H-NbS$_2$. Our results show that while both orbital and spin Nernst conductivities in MoS$_2$ require electron or hole doping, both effects are intrinsically present in metallic NbS$_2$. Our work reveals the central role of orbital and spin Berry curvatures, identifies doping as an effective route for tuning orbital and spin Nernst responses, and proposes a possible experimental setup for detecting these effects in monolayer transition metal dichalcogenides.  %
\end{abstract}

\maketitle

\section{Introduction}
The discovery of Berry phase \cite{berry1,berry2} has significantly broadened our understanding of charge, spin, and orbital transport phenomena \cite{berry3,berry4,berry5,berry6,berry7,normalspinhall,orbitronics_1,orbitronics_2}. Among these, the Hall \cite{Hall1880,Hall1881,quantum_hall1} and Nernst effects \cite{Nernst1886} constitute two fundamental mechanisms through which transverse current arises in response to an external field or a temperature gradient. While the ordinary Hall and Nernst effects have been studied for decades and remain central to semiconductor physics and thermoelectric applications, their analogues involving the spin degrees of freedom \cite{normalspinhall,normalspinnernst,spintronics1,spintronics2}, rather than the charge degrees of freedom of electrons, form the foundation of low-power, high-frequency, and energy-efficient next-generation quantum devices \cite{quantum_devices1,quantum_devices2,quantum_devices3,experiment_1,orbitronics_3,quantum_orbital_magnetization,pure_spin_current}.

The spin Hall and spin Nernst effects,  where a spin current is generated in response to an applied electric field and a temperature gradient, respectively, have been extensively investigated in spin–orbit–coupled materials \cite{spin_hall_nernst,spin_hall_new,intrinsic_spin_hall,giant_spin_hall,strong_intrinsic_spin_hall}. However, their inherent dependence on spin–orbit coupling (SOC) limits the magnitude of these effects due to the relatively small SOC energy scale, restricting their applicability primarily to materials containing heavy elements \cite{experiment_3,heavy_spin_hall2}. Recent studies \cite{Go2018,Canonico2020,new_orbitalhall3,greater_orbital_hall,Cysne2021,large_ohe,gapped_graphene,new_orbitalhall2,Sala2023,Sun2024,Salemi2019,Bhowal2020,Dutta2026} have shown that orbital-driven effects, rather than their spin counterparts, are particularly promising in this context. The orbital-driven phenomena do not require SOC, and, consequently, they can exhibit much larger orbital conductivities, leading to a significantly broader range of relevant materials. Beyond potential applications, these orbital transport phenomena also open a fundamentally new paradigm in condensed matter physics, challenging the long-standing belief that orbital degrees of freedom are quenched in solids \cite{Kittel2004,Cysne2025}.

\begin{figure}[t]
    \centering
   
\includegraphics[width = \columnwidth]{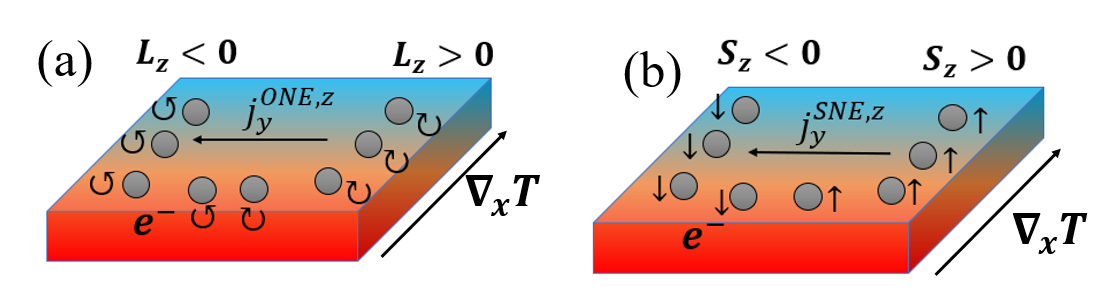}
\caption{(a) Schematic depiction of the (a) orbital Nernst effect and (b) spin Nernst effect. The red and blue regions denote regions of higher and lower temperatures. The grey circles denote electrons of different orbital ($L_z$) and spin angular moments ($S_z$).}
    \label{fig1}
\end{figure}

Among these orbital transport processes, the orbital Hall effect (OHE) is especially noteworthy and has been investigated in both inversion-symmetric and inversion-broken systems~\cite{Go2018,new_orbitalhall3,Canonico2020,Tarik2021,gapped_graphene, Tarik2022,verylargeorbital3,large_ohe,gigantic,ohe,centrosymmetric1}. Following early theoretical predictions, experimental signatures of the OHE have now been reported \cite{Sala2022,orbital_hall_experiment1,orbital_hall_experiment2,OrbitalTransport,orbital_transport2,orbital_transport_3}. In contrast, its thermal analogue, the orbital Nernst effect (ONE), in which a transverse orbital current is generated in response to a temperature gradient, remains comparatively underexplored. To the best of our knowledge, it has so far been proposed only in inversion-symmetric systems \cite{one}. It is important to emphasize, however, that the ONE is one of the most promising orbitronic effects in the sense that it not only enables SOC-free orbitronic devices analogous to those based on the OHE but also offers the key advantage of energy harvesting by converting thermal energy into orbital current.

In the present work, we propose the existence of the ONE and the spin Nernst effect (SNE) in monolayers of transition metal dichalcogenides (TMDCs) with the general formula $M$X$_2$ in their 2H phase. Monolayers of TMDCs have long been at the center of attention due to their valley-contrasting physics, strong spin–orbit coupling, and tunable band structures \cite{tmdc1,tmdc2}. The existence of the OHE and the spin Hall effect (SHE) in this family of materials has also been studied. Our work demonstrates the presence of the corresponding Nernst effects in these materials, highlights their characteristic features, and proposes a possible setup for the experimental detection of the predicted effects.

We show that, in contrast to the OHE, the ONE vanishes for insulating $M$X$_2$ systems. However, with electron and/or hole doping, the ONE can be induced even in the insulating case, while in metallic systems it appears intrinsically. Similar to the OHE, the ONE exists without SOC. In contrast, both the SHE and SNE depend strongly on the presence of SOC and vanish in its absence. For hole-doped systems, the physics of the ONE and SNE is dominated by the orbital and spin Berry curvature contributions near the valley points, and accordingly, the valley model captures the essential physics and provides crucial insights in this regime. In contrast, for electron-doped systems, the physics of the ONE and SNE is dictated by the orbital and spin Berry curvatures around the $\Gamma$ point.

Our findings are based on analytical results obtained from the valley model as well as a transition-metal $d$-orbital tight-binding Hamiltonian with parameters specific to two representative monolayers of $M$X$_2$, viz., MoS$_2$ and NbS$_2$, where the former is insulating and the latter metallic. Our work extends the study of the ONE from centrosymmetric systems to noncentrosymmetric systems and identifies the family of TMDCs as promising candidate materials for observing this effect. The proposed ONE and SNE hold promise for energy harvesting within the framework of orbitronics.

The rest of the manuscript is organized as follows. We begin with the structural details of monolayer TMDCs in section~\ref{sec2}, followed by the analytical results of our valley model in section~\ref{sec3}, which provide crucial insight into the role of different parameters, including hole doping, in the ONE and SNE. In section~\ref{sec4}, we present the full Brillouin zone (BZ) results for the ONE and SNE based on tight-binding models specific to MoS$_2$ and NbS$_2$, followed by our proposal of an experimental setup for detecting the predicted effects in section~\ref{sec5}. Finally, in section~\ref{sec6}, we summarize our findings, and outline open questions for future investigations.

\begin{figure}[t]
\centering
\includegraphics[width=\columnwidth]{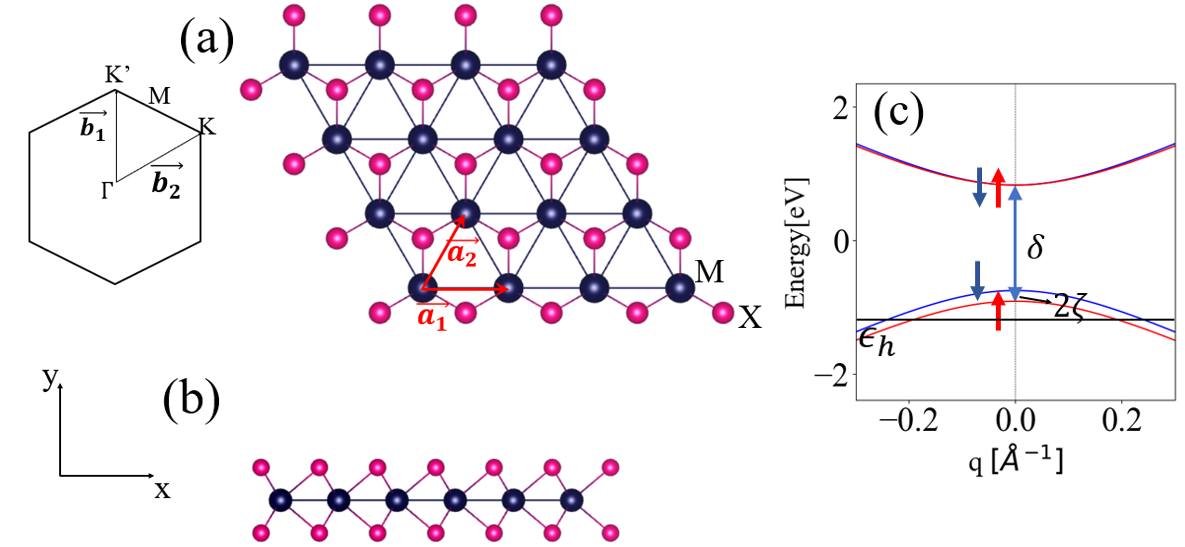}
\caption{ Crystal structure of monolayer transition metal dichalcogenide in its 2H phase.
    (a) Top and 
    (b) side view of the monolayer of transition metal dichalcogenide, MX$_2$.
    Here, M denotes the transition metal atom, and X denotes the chalcogen atom. 
    The position of the chalcogen X atom above and below the the plane of transition metal M atoms breaks the inversion symmetry.
    The structure overall can be seen as a quasi-2D structure. The inset of (a) shows the 2D BZ. (c) Schematic band structure of the valley model around the K' valley point. The blue and red colors denote the down and up-spin polarized bands, respectively.
    The black horizontal line denotes the Fermi Energy ($\epsilon_h$). The considered parameters of the valley model are $\delta$ = 1.66 eV, t = 1.22 eV, and $\zeta$ = 0.08 eV.}
\label{fig2}
\end{figure}

\section{Crystal structure}\label{sec2}
The monolayers of transition metal dichalcogenides (TMDCs) have a generic formula, \(MX_2\), where $M$ denotes the transition metal atom and $X$ denotes the chalcogen atom. In the present work, we have studied the 2H phase of two different monolayers of TMDCs, MoS$_2$ and NbS$_2$, which are representatives of insulating and metallic TMDCs. Both monolayers crystallize in the $\cal I$ symmetry-broken $D_{3h}$ point group symmetry. The crystal structure of the monolayer of TMDC is shown in Fig. \ref{fig2}a. The unit cell contains one formula unit, i.e., one $M$ and two $X$ atoms. The $M$ atoms form a triangular lattice, dictated by the translation vectors,
\begin{equation}\label{eq:lattice vector}
\vec{a}_1 = a \hat{x}, \hspace{3mm}
\vec{a}_2 = a(-\frac{1}{2}\hat{x} + \frac{\sqrt{3}}{2}\hat{y}).
\end{equation}
Each $M$ atom is surrounded by six $X$ atoms, forming a $MX_6$ trigonal prismatic environment. As shown in Fig. \ref{fig2}b, the two chalcogen $X$ atoms are sitting in the out-of-plane direction, such that the top and bottom $X$ atoms preserve the mirror symmetry but break the $\cal I$ symmetry. Both these materials are nonmagnetic, thereby preserving the time-reversal ($\cal T$) symmetry. 
The corresponding BZ (see the inset of Fig. \ref{fig2}a) of the $MX_2$ structure is dictated by the following reciprocal lattice vectors, obtained from Eq. (\ref{eq:lattice vector}),
\begin{align}\label{eq:reciprocal vector}
 \vec{b}_1 &= \frac{2\pi}{\sqrt{3}a}(\sqrt{3}\hat{x} + \hat{y}), \hspace{3mm} \vec{b}_2 = \frac{4\pi}{\sqrt{3}a}\hat{y}.
\end{align}
\section{Results of Valley Model}\label{sec3}

\subsection{Model Hamiltonian}
We start with a low-energy model in the vicinity of the \(K\) and \(K'\) points, also known as valley points, in the BZ of the monolayer of $MX_2$. The valley model is given by \cite{large_ohe}, 

\begin{equation}\label{valleymodel}
\mathcal{H}(\vec{q}) = (\vec{d} \cdot \vec{\tau}) \otimes {\rm I}_s + \frac{\nu \zeta}{2} \left( \tau_z + 1 \right) \otimes \sigma_z.
\end{equation}
Here, $\vec q$ is defined with respect to the valley points ($K, K'$), viz., \(\vec{q} = \vec{k} - \vec{K}\) or \(\vec{q} = \vec{k} - \vec{K'}\) and $\nu$ is the valley index, with \(\nu = \pm 1\) corresponding to the \(K\) and \(K'\) valley points, respectively. \(\vec{\sigma}\) and \(\vec{\tau}\)  are respectively the Pauli Matrices in the spin basis $\uparrow, {\rm and} \downarrow$, and the pseudo-spin basis of the two transition metal $d$ orbitals, \(|v\rangle = \frac{1}{\sqrt{2}}(|x^2 - y^2\rangle + i \nu |xy\rangle)\) and \(|c\rangle = |3z^2 - r^2\rangle\). ${\rm I}_s$ is the identity operator in the spin basis and $\zeta$ is the spin-orbit coupling strength. $\vec d$ is defined by the electronic hopping parameter $t$ and the energy gap $\delta$, where \(d_x = \nu t q_x a\), \(d_y = -t q_y a\), and \(d_z = -\delta / 2\). Here, \(a\) is the lattice constant.\\
The diagonalization of the Hamiltonian in Eq.~(\ref{valleymodel}) gives us the following energy eigenvalues,
\begin{equation}\label{valley_energy}
\epsilon_s^\pm = \frac{1}{2}\left\{ \nu s \zeta \pm \left[ (\delta - \nu s \zeta)^2 + 4 t^2 a^2 q^2 \right]^{1/2} \right\},
\end{equation}
where \( s = \pm 1 \) denotes the $\uparrow$ and $\downarrow$ spin-polarization of the conduction or valence bands, represented by the superscript \( \pm \). 
The spin-polarization of the bands around one of the valley points, and the SOC-induced energy splitting 2\(\zeta\) between the bands
are shown in Fig. \ref{fig3}a.

The corresponding energy eigenfunctions in the basis set (\( |v \uparrow\rangle, |c \uparrow\rangle, |v \downarrow\rangle, \) \( |c \downarrow\rangle \)) are given by,

\[|\psi_{s=1}^\pm(\vec{q})\rangle = \mathcal{N}_\pm
\begin{bmatrix}
1  \\
\mathcal{C}_{\pm} \\
0 \\
0 \\
\end{bmatrix},  \hspace{3mm}|\psi_{s=-1}^\pm(\vec{q})\rangle = \mathcal{N}_\pm
\begin{bmatrix}
0  \\
0 \\
1 \\
\mathcal{C}_{\pm}\\
\end{bmatrix} .
 \]
Here $\mathcal{C}_\pm = \frac{B^s\mp\sqrt{(B^s)^2+b^2}}{b^s}$ with  $B^s = \frac{\delta - \nu s\zeta}{2}, \quad b^s = ta (\nu q_x \pm i s q_y), \quad b^2 = t^2 a^2 (q_x^2 + q_y^2) $, and \( \mathcal{N_\pm}\) being the corresponding normalization constants.  
\subsection{Methods for Calculating the Orbital and Spin Nernst Conductivity}

The orbital and spin Nernst effects refer to the generation of a transverse flow of orbital and spin moments, called the orbital and spin currents, respectively, in response to an applied temperature gradient, viz.,
\begin{equation}\label{one_current}
j_{\alpha}^{\text{ONE},\gamma} = \sigma^{\gamma,\text{ONE}}_{\alpha\beta} \frac{\partial T}{\partial \beta},
\end{equation}
\begin{equation}\label{sne_current}
j_{\alpha}^{\text{SNE},\gamma} = \sigma^{\gamma,\text{SNE}}_{\alpha\beta} \frac{\partial T}{\partial \beta}.
\end{equation}
Here, $j_{\alpha}^{\text{ONE},\gamma}$ ($j_{\alpha}^{\text{SNE},\gamma}$) is the orbital (spin) Nernst current density in the $\alpha$ direction with the orbital (spin) angular momentum polarized along the $\gamma$ direction in response to the temperature gradient $\frac{\partial T}{\partial \beta}$ along the $\beta$ direction.\\
ONE and SNE are the thermal counterparts of the orbital and spin Hall effects, respectively, in which a transverse orbital or spin current is generated in response to an applied electric field. Consequently, the orbital and spin Nernst conductivities, $\sigma^{\gamma,\text{ONE}}_{\alpha\beta}$ and $\sigma^{\gamma,\text{SNE}}_{\alpha\beta}$, can be computed from the energy derivative of the orbital and spin Hall conductivities, $\sigma^{\gamma,\text{OHE}}_{\alpha\beta}$ and $\sigma^{\gamma,\text{SHE}}_{\alpha\beta}$, respectively~\cite{Culter1969,one},
\begin{equation}\label{one_conductivity}
    \sigma^{\gamma,\text{ONE}}_{\alpha\beta} \equiv \frac{\pi^2 k^2_B T}{-3e}(\frac{d}{d\epsilon} \sigma^{\gamma,\text{OHE}}_{\alpha\beta})\Bigg|_{\epsilon=\epsilon_F}~{\rm and}
\end{equation}
\begin{equation}\label{sne_conductivity}
   \sigma^{\gamma,\text{SNE}}_{\alpha\beta} \equiv \frac{\pi^2 k^2_B T}{-3e}(\frac{d}{d\epsilon} \sigma^{\gamma,\text{SHE}}_{\alpha\beta})\Bigg|_{\epsilon=\epsilon_F}.
\end{equation}
Here, $\epsilon_{F}, T, k_B, {\rm and}~ e$ are respectively the Fermi Energy, the temperature, the Boltzmann Constant, and the electronic charge.
The OHC, $\sigma^{\gamma,\text{OHE}}_{\alpha\beta}$, in Eq. (\ref{one_conductivity}) can be computed by summing the orbital Berry curvature, $\Omega^{\gamma,\text{orb}}_{n,\alpha\beta}(\vec{q})$, over the occupied part of the BZ \cite{large_ohe},

\begin{equation}\label{ohe_conductivity}
    \sigma^{\gamma,\text{OHE}}_{\alpha\beta} = - \frac{e}{N_k A_c} \sum_{n \vec{q}}^\text{occ} \Omega^{\gamma,\text{orb}}_{n,\alpha\beta}(\vec{q})
\end{equation}
Similar to the OHC, the SHC can also be written as a BZ sum of the spin Berry curvature $\Omega^{\gamma,\text{spin}}_{n,\alpha\beta}(\vec{q})$, viz.,   
\begin{equation}\label{she_conductivity}
    \sigma^{\gamma,\text{SHE}}_{\alpha\beta} = - \frac{e}{N_k A_c} \sum_{n \vec{q}}^\text{occ} \Omega^{\gamma,\text{spin}}_{n,\alpha\beta}(\vec{q}).
\end{equation}
Here, $n, N_k, {\rm and}~ A_c$ are respectively the band index, total number of $k$ points in the BZ, and the area of the monolayer. The orbital Berry curvature \(\Omega_{n,\alpha\beta}^{\gamma,\text{orb}}(\vec{q})\) and the spin Berry curvature $\Omega^{\gamma,\text{spin}}_{n,\alpha\beta}(\vec{q})$ of the $n$th band in the equation above can be computed from the energy eigenvalues and eigenvectors of the Hamiltonian ${\cal H}(\vec q)$ using the Kubo formula \cite{large_ohe},
\begin{align} 
\Omega^{\gamma,\text{orb}}_{n,\alpha\beta}(\vec{q}) = 2\hbar\sum_{n' \neq n} 
\text{Im} \bigg[
\frac{
\langle n(\vec{q}) | \mathcal{J}_{\alpha}^{\gamma,\text{orb}} | n'(\vec{q}) \rangle
\langle n'(\vec{q}) | v_\beta | n(\vec{q}) \rangle
}{
(\epsilon_{n'} - \epsilon_n)^2
}
\bigg] \label{obc}
\end{align}
and,
\begin{align}
\Omega^{\gamma,\text{spin}}_{n,\alpha\beta}(\vec{q}) = 2\hbar\sum_{n' \neq n} 
\text{Im} \bigg[
\frac{
\langle n(\vec{q}) | \mathcal{J}_{\alpha}^{\gamma,\text{spin}} | n'(\vec{q}) \rangle
\langle n'(\vec{q}) | v_\beta | n(\vec{q}) \rangle
}{
(\epsilon_{n'} - \epsilon_n)^2
}
\bigg]. \label{sbc}
\end{align}
\noindent 

Here, the orbital current operator, 
$
\mathcal{J}_\alpha^{\gamma,\text{orb}}  = \frac{1}{2} \left\{ v_\alpha, L_\gamma \right\},
$ and the spin current operator, 
$\mathcal{J}_\alpha^{\gamma,\text{spin}}  = \frac{1}{4} \left\{ v_\alpha, s_\gamma \right\}$, with the velocity operator 
$
v_\alpha = \frac{1}{\hbar} \frac{\partial H}{\partial q_\alpha}$, and $\frac{1}{2}s_\gamma$ and $L_\gamma$ are respectively the spin and orbital angular momentum operators. The \(|n(\vec{q})\rangle\) and \(\epsilon_n\) are the $n$th eigenvector and the corresponding energy eigenvalue of the Hamiltonian. Thus, the orbital Nernst conductivity (ONC) for the Hamiltonian in Eq. (\ref{valleymodel}) can be computed using Eqs. (\ref{one_conductivity}), (\ref{ohe_conductivity}), and (\ref{obc}). Similarly, the spin Nernst conductivity (SNC) can be computed using Eqs. (\ref{sne_conductivity}), (\ref{she_conductivity}), and (\ref{sbc}).

\subsection{Results of  Orbital and Spin Nernst Effects}

We now proceed to obtain the analytical expression of the ONC for the valley model. We begin with orbital Berry curvature, which, in the present case, has only two non-zero components, $\Omega_{n,xy}^{z,\text{orb}}(\vec{q}) \quad \text{and} \quad \Omega_{n,yx}^{z,\text{orb}}(\vec{q})$, as follows from Eq. (\ref{obc}) and $v_z = 0$. 

Using Eq. (\ref{obc}), we obtain the orbital Berry curvature of the valence band of the valley model,
\begin{equation}\label{valley_obc}
\Omega_{n,yx}^{z,\text{orb}}(\vec{q}) = -\Omega_{n,xy}^{z,\text{orb}}(\vec{q})=\frac{2t^2 a^2 (\delta -  \nu s \zeta)}{[(\delta -  \nu s \zeta)^2 + 4t^2 a^2 q^2]^\frac{3}{2}}.
\end{equation}
As evident from the analytical expression, the orbital Berry curvature can exist even without the SOC. Furthermore, the orbital Berry curvature has the same sign at the two valley points, with $\Omega_{n,yx}^{z,\text{orb}}\bigg|_{q=0} = \frac{2 t^2 a^2}{\delta^2}$ for $\zeta =0 $, resulting in a net non-zero OHC, when summed over the occupied part of the BZ. 

To compute the OHC, we integrate the orbital Berry curvature over the occupied part of the BZ. For this, we assume a metallic band structure with the Fermi energy, $\epsilon_h$, lying on the spin-polarized valence bands, as shown in Fig. \ref{fig2}c. This leads to hole pockets around the valley points at $\vec q=0$. We note that the constant energy contours, in this case, are circular in shape as evident from Eq. (\ref{valley_energy}). The radii of these circular hole pockets around the valley point are determined by the hole concentration $n_h$ and the spin-orbit splitting of the spin-polarized bands. 

Since the OHC exists even without $\zeta$, for simplicity, here we consider the $\zeta=0$ case. In this case, therefore, we have degenerate bands and the hole pocket radius is determined solely by $n_h$, where $0 \le n_h \le 2$. Since there are only two valence bands for the valley model, $n_h=0, {\rm and}$~ 2 correspond to the completely filled and empty bands, respectively. Let us assume that the Fermi momentum corresponding to $n_h=0$ is \(q_c\), whereas that for a non-zero value of $n_h$ is $q_h$. We note that $n_h=0$ represents an insulating band structure, while $0 < n_h < 2$ corresponds to a metallic system, relevant to insulating MoS$_2$ and metallic NbS$_2$ monolayers, as we discuss later. 

Approximating the area of the two circular regions of radius $q_h$ around the two valley points to be equal to the occupied part of the BZ area, we get an estimate for $q_h$, viz.,
\begin{equation}\label{BZ_eq}
    2\pi q_h^2 = \frac{n_h}{2} A_\text{BZ} = \frac{n_h}{2} \frac{8\pi^2}{\sqrt{3}a^2}.
\end{equation}
Here, $A_\text{BZ}=\frac{8\pi^2}{\sqrt{3}a^2}$ is the BZ area [see Eq. (\ref{eq:reciprocal vector})]. For $n_h=0$, we get $2\pi q_c^2 = \frac{8\pi^2}{\sqrt{3}a^2}$.

Consequently, we can compute the OHC for $0 < n_h < 2$ by integrating the orbital Berry curvature over an annular region of inner radius $q_h$ and outer radius $q_c$. It is, therefore, given by,
\begin{align}\label{valley_ohe}
\sigma^{z, \text{OHE}}_{yx} 
&= - \frac{2e}{(2\pi)^2} \sum_{s=\pm1} \int_{q_h}^{q_c} 2\pi q~ \Omega^{z,\text{orb}}_{s, yx} (\vec{q})  dq \, \notag \\
&= -\frac{e}{\pi} \left[ 
\frac{\delta}{\sqrt{\delta^2 + 4t^2a^2q_h^2}} 
- \frac{\delta}{\sqrt{\delta^2 + \frac{16\pi t^2}{\sqrt{3}}}} 
\right]. 
\end{align}
The factor of 2 on the right-hand side accounts for the two valleys, for which the orbital Berry curvature contributions are identical in the absence of $\zeta$.\\
Rewriting the OHC in terms of the Fermi energy $\epsilon_h = -\frac{1}{2}\sqrt{\delta^2 + 4t^2a^2q_h^2}$, we get,
\begin{equation}\label{valley_ohe_eh}
\sigma^{z, \text{OHE}}_{yx} \approx \frac{e}{\pi} \left[ \frac{\delta}{2\epsilon_h} + \xi \right], {\rm where~} \xi = \frac{\delta}{\sqrt{\delta^2 + \frac{16\pi t^2}{\sqrt{3}}}}.
\end{equation}
It is clear from the above equation that for $n_h=0$ (i.e., $q_h=0$), both bands are fully occupied, such that $\epsilon_h=-\delta/2$, and, consequently, $\sigma^{z, \text{OHE}}_{yx}|_{n_h=0}=-\frac{e}{\pi} (1-\xi)$. Since $\xi <1$, this corresponds to the maximum value of the OHC. With the increase in $n_h$, the magnitude of $\epsilon_h$ increases, or in other words, the Fermi energy goes deeper in the valence bands, leading to a reduction in the OHC. 

Following Eq. (\ref{valley_ohe_eh}), we calculate our desired quantity, the ONC, using Eq. (\ref{one_conductivity}). We note that the ONC vanishes for $n_h=0$, as directly follows from the energy derivative of $\sigma^{z, \text{OHE}}_{yx}|_{n_h=0}$. Thus, the ONE is absent in insulating systems. However, in the presence of doping, we can 
induce ONE, and the magnitude of the ONC is given by,
\begin{equation}\label{valley_one}
  \sigma^{z, \text{ONE}}_{yx} = \frac{\pi \delta {k^2_B} T}{6{\epsilon_h}^2}.  
\end{equation}
We note that the analytical expression obtained for the ONC exhibits a stronger dependence on the Fermi energy compared to its electrical counterpart, as shown in Eq. (\ref{valley_ohe_eh}).

We now move on to the discussion of the SNE. Similarly to the orbital Berry curvature, the spin Berry curvature also has only two non-zero components, 
$\Omega_{n,xy}^{z,\text{spin}}(\vec{q}) \quad \text{and} \quad \Omega_{n,yx}^{z,\text{spin}}(\vec{q})$. The value of these components of the spin Berry curvature is calculated using Eq. (\ref{sbc}). For this, we use the following form of the $s_z$ operator in the basis set \( (|v \uparrow\rangle, |c \uparrow\rangle, |v\downarrow\rangle, |c \downarrow\rangle) \) of the Hamiltonian,
\[
s_z = \hbar(\sigma_z \otimes \mathcal{I})= \hbar
\begin{bmatrix}
1 & 0 & 0 & 0 \\
0 & 1 & 0 & 0 \\
0 & 0 & -1 & 0 \\
0 & 0 & 0 & -1 \\
\end{bmatrix},
\]
where \(\sigma_z\) is the usual Pauli spin matrix and \(\mathcal{I}\) is the identity operator. The calculated spin Berry curvature for the valence band of the valley model is given by, 
\begin{equation}\label{valley_sbc}
\Omega_{n,yx}^{z,\text{spin}}(\vec{q}) = \frac{ \nu s t^2 a^2 (\delta -    \nu s \zeta)}{[(\delta -   \nu s \zeta)^2 + 4t^2 a^2 q^2]^\frac{3}{2}}. 
\end{equation}
We note that in the absence of \(\zeta\), the spin Berry curvatures for the two oppositely spin-polarized ($\nu=\pm 1$) valence bands are identical in magnitude and opposite in sign. As a result, their contributions cancel each other, leading to vanishing SHC in the absence of \(\zeta\). 
In the presence of the spin-orbit interaction, however, the two contributions differ in magnitude, leading to a tiny but non-zero value of SHC. 

The sum of spin Berry curvature over the occupied part of the BZ, using Eq. (\ref{she_conductivity}), gives us the SHC of the valley model. Similarly to the OHC, we, therefore, convert the summation to an integral, and the limits of the integral are determined by the hole concentration $n_h$, as discussed earlier. The resulting SHC in the presence of \(\zeta\), therefore, becomes,  
\begin{widetext}
    \begin{align}
\sigma^{z, \text{SHE}}_{yx} 
&= - \frac{2e}{(2\pi)^2} \sum_{s=\pm1} \int_{q_h}^{q_c} 2\pi q~ \Omega^{z,\text{spin}}_{s, yx} (\vec{q})  dq \, \notag 
 = -\frac{et^2 a^2}{\pi} \int_{q_h}^{q_c} q\, dq \left[
\frac{\delta - \zeta}{\left[ (\delta - \zeta)^2 + 4t^2 a^2 q^2 \right]^{3/2}} 
\right.
 \left.
 - \frac{\delta + \zeta}{\left[ (\delta + \zeta)^2 + 4t^2 a^2 q^2 \right]^{3/2}} 
\right]\\ \nonumber
&= -\frac{e}{4\pi} \left[
\frac{1}{\sqrt{1 + \xi_{+} q_c^2}} 
- \frac{1}{\sqrt{1 + \xi_- q_c^2}} 
+ \frac{1}{\sqrt{1 + \xi_- q_h^2}} 
- \frac{1}{\sqrt{1 + \xi_+ q_h^2}} 
\right] \\
&= \sigma^{z, \text{SHE}}_{yx,0} 
+ \frac{e}{4\pi} \left[
 \frac{\delta-\zeta}{2\epsilon_h-\zeta} 
- \frac{\delta+\zeta}{2\epsilon_h+\zeta}
\right]. \label{valley_she_eh}
\end{align}
 
\end{widetext}
Here, \(\xi_\pm = 4t^2a^2/(\delta \pm \zeta)^2\) and $\sigma^{z, \text{SHE}}_{yx,0}$ is the SHC in the absence of doping ($n_h=0$), i.e.,
\begin{equation}\label{valley_she_undoped}
 \sigma^{z, \text{SHE}}_{yx,0} = -\frac{e}{4\pi} \left[
\frac{1}{\sqrt{1 + \xi_+ q_c^2}} 
- \frac{1}{\sqrt{1 + \xi_- q_c^2}} 
\right]. 
\end{equation}
In obtaining the last equality in Eq. (\ref{valley_she_eh}) we use the expression for the Fermi energy, $\epsilon_h = \frac{1}{2}\left[ \nu s \zeta - \left[ (\delta - \nu s \zeta)^2 + 4 t^2 a^2 q_h^2 \right]^{1/2}\right]$.
As follows from Eq. (\ref{valley_she_eh}), the SHC for the doped system vanishes for $\zeta=0$, similarly to the undoped case. Furthermore, in the limit, $\zeta << \delta$, we find $\sigma^{z,\text{SHE}}_{yx} \propto \zeta$, highlighting the strong dependence of the SHE on the SOC strength.

We next calculate the SNC using Eq. (\ref{sne_conductivity}). For the undoped case, the SNC vanishes, as directly follows from Eq. (\ref{valley_she_undoped}). In the presence of doping, however, the SNC is non-zero, and using Eq. (\ref{valley_she_eh}) we find the corresponding value,
\begin{align}
  \sigma^{z, \text{SNE}}_{yx} \notag
&= \frac{\pi {k^2_B} T}{6}\Bigg[
 \frac{\delta-\zeta}{(2\epsilon_h-\zeta)^2} - 
 \frac{\delta+\zeta}{(2\epsilon_h+\zeta)^2}
\Bigg].\label{valley_sne}   
\end{align}  
For a given doping concentration and for weak SOC strength $\zeta$, the SNC reduces to,
$ \sigma^{z, \text{SNE}}_{yx} = -\frac{\pi {k^2_B} T\zeta}{12\epsilon_h^2}\left(
 1-\frac{\delta}{\epsilon_h}
\right) + O (\zeta^2). 
$
This indicates that the SNC is also proportional to the SOC strength $\zeta$, similarly to the SHC. We also note that the SNC has a stronger dependence on the Fermi energy compared to the SHC, similarly to its orbital counterparts, as discussed above. 

\section{Full Brillouin Zone Result}\label{sec4}
The simple valley model, as described above, provides crucial insight into both orbital and spin Nernst effects. However, since the corresponding Hall and Nernst conductivities are dictated by the contributions coming from the entire BZ, a full BZ study is required to capture the realistic picture, in which there might be important contributions other than the valley points. Motivated by this, here, we study the tight-binding model for the transition metal $d$ bands, which provides the platform to investigate the entire BZ for the study of ONC and SNC.\\

\subsection{Tight Binding Model}\label{tba}

We begin by constructing a tight-binding model on the triangular lattice, relevant to $MX_2$ (see Fig. \ref{fig2}a),  
\begin{eqnarray}\label{TBM-BZ}\nonumber
 \hat{H} &= & \sum_{i,m,m',\sigma}\epsilon_{imm'} \hat{c}_{im\sigma}^\dagger \hat{c}_{im'\sigma}\\  &+ &\sum_{ i,j,m,m',\sigma}t_{ij}^{mm'} \hat{c}_{im\sigma}^\dagger \hat{c}_{jm'\sigma}+\zeta \vec L \cdot \vec S.   
\end{eqnarray}

Here, $i,j$ denote the transition metal $M$ site index, $m,m'$ refer to the five $M$-$d$ orbitals, \(|d_{xy}\rangle,|d_{yz}\rangle , |d_{3z^2-1}\rangle, |d_{xz}\rangle, {\rm and}~|d_{x^2-y^2}\rangle\), and $\sigma$ is the spin index. $\epsilon_{imm'}$ and $t_{ij}^{mm'}$ are respectively the onsite energy of the $i$th site and the electronic hopping between the neighboring $m$ orbital at the $i$th site and $m'$ orbital at the $j$th site of the $M$ ion. 

\begin{figure*}[t]
    \centering
    \includegraphics[scale=0.56]{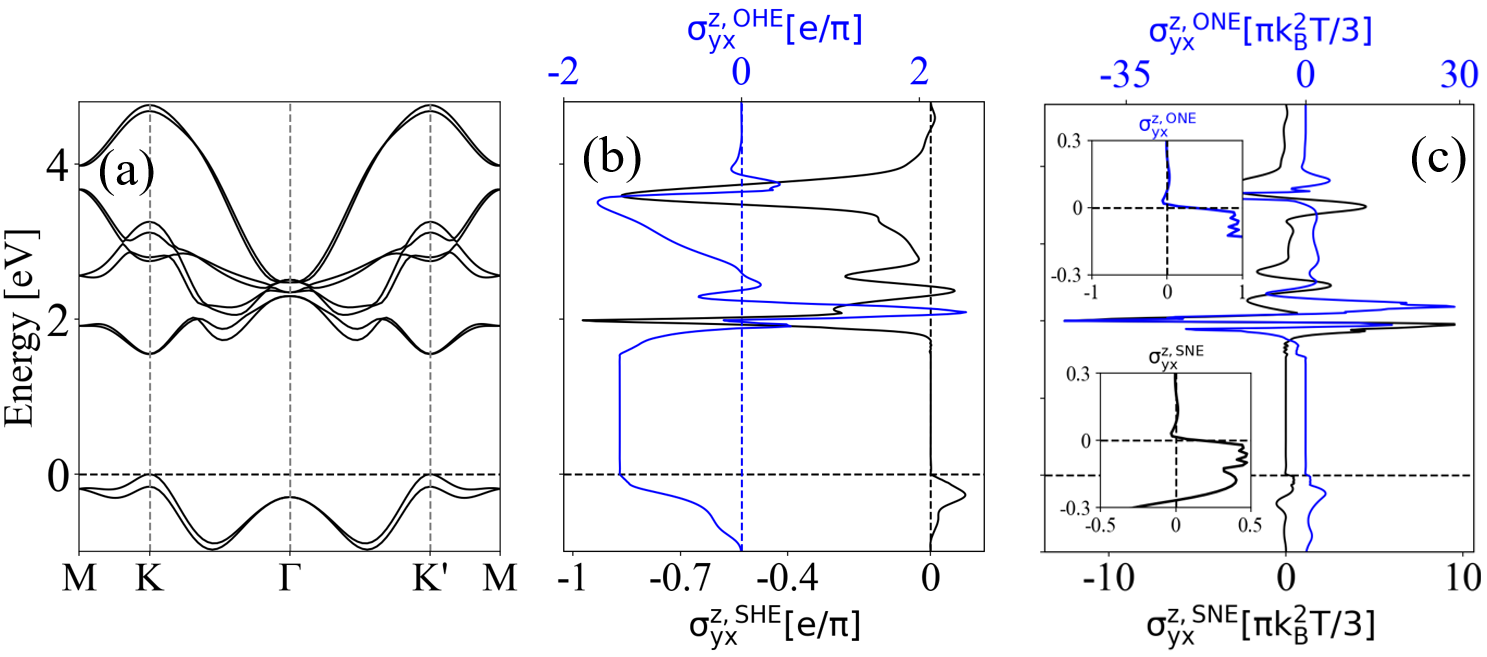}
    \caption{Orbital and spin Hall and Nernst conductivities in MoS$_2$.
    (a)  Band structure of the MoS$_2$ tight-binding model in the presence of SOC. (b) The corresponding variations of OHC (blue) and SHC (black) as a function of energy. Here, the top x-axis is for OHC, and the bottom x-axis is for SHC.
    (c) Similar variations for ONC (blue, top x-axis) and SNC (black, bottom x-axis). The insets in (c) show the enlarged view of ONC and SHC near the Fermi energy. The black dashed line indicates the Fermi energy.
    }
    \label{fig3}
\end{figure*}
\begin{figure*}[t]
    \centering
    \includegraphics[scale=0.58]{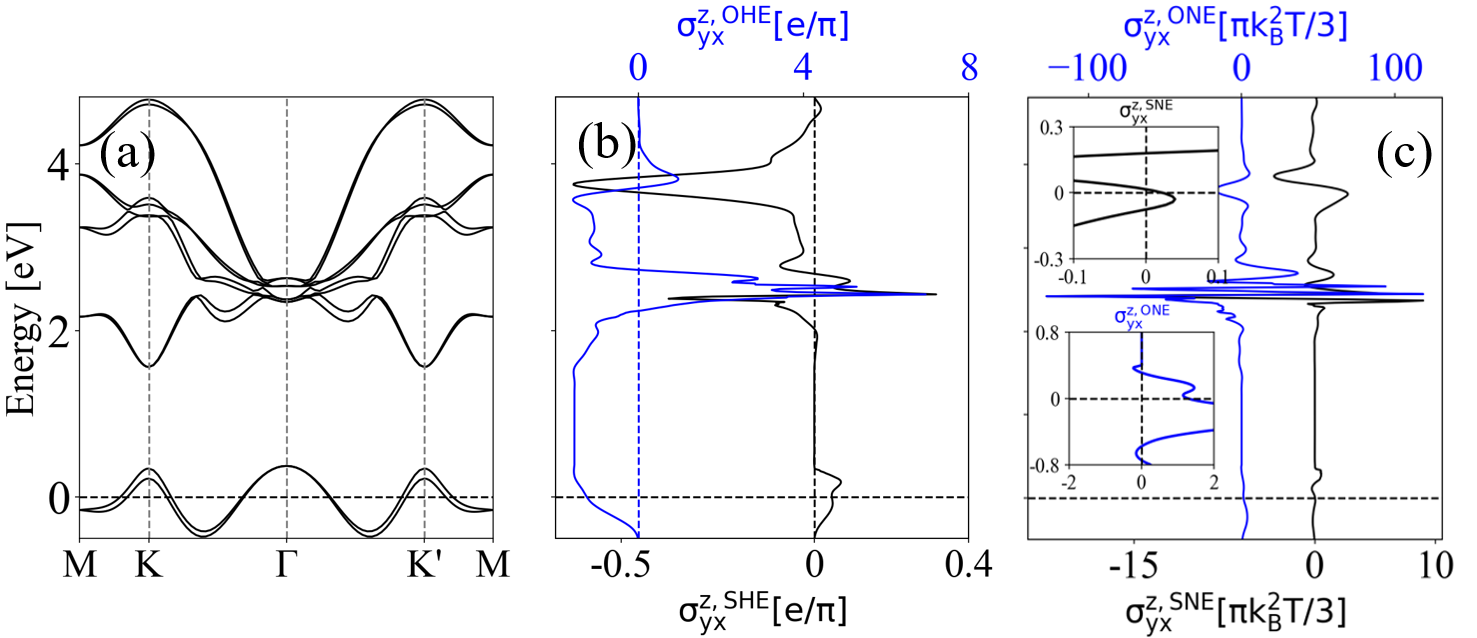}
    \caption{Orbital and spin Hall and Nernst conductivities in NbS$_2$. (a) The band structure of NbS$_2$ in the presence of spin-orbit interaction. The black horizontal dashed line indicates the Fermi energy. 
    (b) The variation of OHC (blue) and SHC (black) in units of \(e/\pi\) as a function of energy. Here, the top and bottom x-axes correspond to the OHC and SHC, respectively.
    (c) The same variation for the ONC (blue, top x axis) and SNC (black, bottom x axis). The insets in (c) show the enlarged view of the ONE and SNE near the Fermi level. 
    }
    \label{fig4}
\end{figure*}

Here, we consider two example materials, MS$_2$ with $ M=$ Mo, Nb, which are representatives of insulating and metallic 2H monolayer TMDCs, respectively. The corresponding parameters of the Hamiltonian, e.g., the onsite energies, and the effective M $d-d$ hopping parameters are extracted using the Nth order muffin-tin orbital (NMTO) method \cite{nmto5}, where we downfold the effect of the S atoms. For both MoS$_2$ and NbS$_2$, we have considered up to the 4th nearest neighbor (NN) hopping in the respective tight-binding models.

The last term in Eq. (\ref{TBM-BZ}) represents the spin-orbit interaction, with $\zeta$, $\vec S$, and $\vec L$ being respectively the SOC constant, spin, and orbital angular momenta in the basis set of the $M-d$ orbitals.
The SOC term in Hamiltonian in the basis set of \(|d_{xy}\uparrow\rangle,\;
|d_{3z^2-1}\uparrow\rangle,\;
|d_{x^2-y^2}\uparrow\rangle,\;
|d_{xz}\uparrow\rangle,\;
|d_{yz}\uparrow\rangle,\;
|d_{xy}\downarrow\rangle,\;
|d_{3z^2-1}\downarrow\rangle,\;
|d_{x^2-y^2}\downarrow\rangle,\;
|d_{xz}\downarrow\rangle,\;
|d_{yz}\downarrow\rangle\) is as follows: 
\begin{widetext}
\[
\mathcal{H}_{\mathrm{SOC}} = \zeta\, \mathbf{L}\cdot\mathbf{S}
= \frac{\zeta}{2}
\begin{bmatrix}
0   & 2i  & 0   & 0   & 0   & 0   & 0   & 0   & -i        & 1 \\
-2i & 0   & 0   & 0   & 0   & 0   & 0   & 0   & 1         & i \\
0   & 0   & 0   & 0   & 0   & 0   & 0   & 0   & -\sqrt{3} & \sqrt{3}i \\
0   & 0   & 0   & 0   & -i  & i   & -1  & \sqrt{3} & 0 & 0 \\
0   & 0   & 0   & i   & 0   & -1  & -i  & -\sqrt{3}i & 0 & 0 \\
0   & 0   & 0   & -i  & -1  & 0   & -2i & 0   & 0 & 0 \\
0   & 0   & 0   & -1  & i   & 2i  & 0   & 0   & 0 & 0 \\
0   & 0   & 0   & \sqrt{3} & \sqrt{3}i & 0 & 0 & 0 & 0 & 0 \\
i   & 1   & -\sqrt{3} & 0 & 0 & 0 & 0 & 0 & 0 & -i \\
1   & -i  & -\sqrt{3}i & 0 & 0 & 0 & 0 & 0 & -i & 0
\end{bmatrix}
\].
\end{widetext}
We transform the Hamiltonian in Eq. (\ref{TBM-BZ}) 
into the Bl\"och function basis and then diagonalize the Hamiltonian $H(\vec k)$, where $\vec k$ is the Bloch momentum, to obtain the band structure in the BZ of the respective materials. 
The computed band structures of MoS$_2$ and NbS$_2$ are shown in Figs. \ref{fig3}a and \ref{fig4}a, respectively. As evident from these figures, one of the key differences between the band structures is that the former is insulating, while the latter is metallic in nature, consistent with the electronic configurations of Mo$^{2+}$: [Kr]$4d^2$ and Nb$^{2+}$:[Kr]$4d^1$, respectively.

\subsection{Results of ONC and SNC}

We now proceed to compute and analyze our results for ONC and SNC in both MoS$_2$ and NbS$_2$ within our constructed 4th-NN tight-binding model. To gain insight into the computed values of ONC and SNC, we further analyze the BZ distribution of the orbital and spin Berry curvature, as we discuss in this section.\\
\subsubsection{Results of MoS$_2$}\label{Results of MoS2}
As discussed in section \ref{sec3}, the key quantities in determining the OHC, ONC, and their spin counterparts, viz., SHC and SNC are respectively the orbital and spin Berry curvatures. We, therefore, compute the distribution of these quantities on the \(k_x-k_y\) plane of the BZ of MoS$_2$, using Eqs. (\ref{obc}) and (\ref{sbc}), respectively, within the tight-binding model description in section \ref{sec4}A. The results of our calculations are shown in Figs. \ref{fig6}a and b. The distribution in these figures corresponds to the Fermi energy, i.e., we sum over the contributions of the two valence bands.

\begin{figure}[t]
    \centering
   \includegraphics[width = \columnwidth]{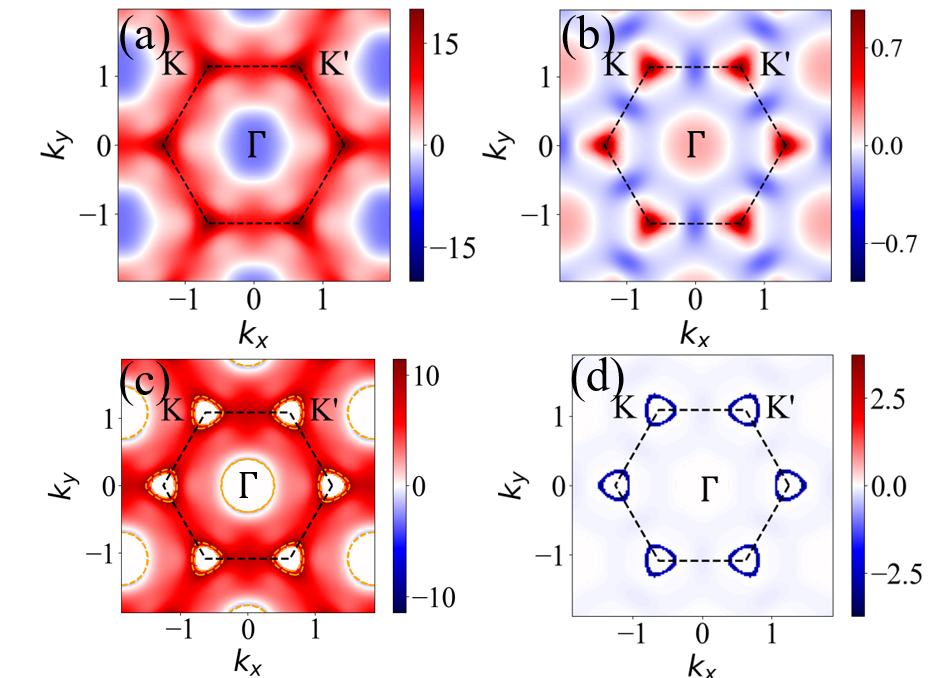}
    \caption{ The $k$ space distribution of
    (a) orbital Berry curvature ($\Omega^{z,\text{orb}}_{n,yx}$) and (b) spin Berry curvature ($\Omega^{z,\text{spin}}_{n,yx}$),  summed over the occupied parts of the BZ of MoS$_2$ on the $k_x-k_y$ plane. 
        Similar distributions of (c) orbital ($\Omega^{z,\text{orb}}_{n,yx}$) and 
        (d) spin Berry curvature ($\Omega^{z,\text{spin}}_{n,yx}$) for NbS$_2$.
        The color maps in (a)-(d) indicate the values (including signs) of  $\Omega^{z,\text{orb}}_{n,yx}$ and $\Omega^{z,\text{spin}}_{n,yx}$ in units of \AA$^2$. 
        The BZ is shown with the black dashed lines and the high-symmetry $k$ points, viz., K, K$'$, and $\Gamma$, are also indicated. The yellow dashed lines in (c) indicate the Fermi surface of NbS$_2$.}
    \label{fig5}
\end{figure}

As seen from Figs. \ref{fig5}a and b, first of all, the orbital Berry curvature has a much higher value than the spin Berry curvature. The predominant contributions of the orbital Berry curvature come from the valley points, and the contributions at the K, K$^\prime$ points are equal in both magnitude and sign. This, in turn, gives us a net OHC when summed over the BZ, as we discuss later. We note that apart from the valley points, as seen in Fig. \ref{fig5}a, the orbital Berry curvature contribution also comes from the region around the $\Gamma$ point. However, the magnitude of $\Omega^{z,\text{orb}}_{n,yx}$ is much weaker around the $\Gamma$ point compared to the valley points. The predominant valley point contributions justify our analysis of the valley model in section \ref{sec3} and are also consistent with our analytical findings in Eq. (\ref{valley_obc}). Similarly to the orbital Berry curvature, the reciprocal space distribution of the spin Berry curvature is also consistent with the presence of time-reversal symmetry in the system. We note that the weak spin Berry curvature contribution can be understood from the near cancellation of the opposite spin Berry curvature contributions from the two valence bands, as we found earlier in Eq. (\ref{valley_sbc}).

To compute the OHC and ONC, we consider a \(100 \times 100\) $k$ mesh and subsequently, the sum of the orbital Berry curvature is carried out to compute the OHC using Eq. (\ref{ohe_conductivity})
 and its energy derivative to compute the ONC using Eq. (\ref{one_conductivity}). The computed OHC and ONC for the tight-binding model relevant to MoS$_2$ are shown in Fig. \ref{fig3}b and c. Interestingly, we find that even though the OHC is non-zero in an insulating system, the value of OHC being constant across the band gap, its energy derivative, which dictates the ONC, vanishes for an insulating system. However, we note that when the Fermi energy lies within the valence or conduction band, relevant to the doped case, the ONC becomes non-zero, consistent with our analytical results in Eq. (\ref{valley_one}).

  Interestingly, we find that the ONC is larger for the electron-doped case than for hole-doped MoS$_2$. The values are particularly high at around 2 eV where multiple band crossings occur at the $\Gamma$ point (see Fig. \ref{fig3}b and c).  Our further analysis of the orbital Berry curvature in this energy regime shows a predominant $\Gamma$ point contribution. Following this, we perform a detailed analysis of the $\Gamma$ point model for the conduction bands, which is discussed in detail in Appendix \ref{app1}. Our analysis points to the important role of the inter-orbital $d_{xz}$-$d_{yz}$ hopping parameters in driving the large orbital Berry curvature in the electron-doped MoS$_2$, which consequently dictates the large values of OHC and ONC.

Similar to the previous case, we also compute the SHC and the SNC using Eqs. (\ref{she_conductivity}) and (\ref{sne_conductivity}) respectively, and the computed results are shown in Figs. \ref{fig3}b and c. In contrast to the OHC, here we find that the SHC is vanishingly small at the Fermi energy, as expected from the corresponding spin Berry curvature distribution. Electron or hole doping, however, leads to a non-zero SHC as evident from Fig. \ref{fig3}b due to incomplete cancellation of the spin Berry curvature. The magnitude of the SHC, however, remains much smaller than the OHC. Similarly to the ONC, the SNC also vanishes at the Fermi energy, while it becomes non-zero in the presence of doping (see Fig. \ref{fig3}c).\\

\subsubsection{Results of NbS$_2$}\label{nbs2}

After discussing the case of insulating MoS$_2$, we now move on to the metallic TMDC NbS$_2$. As we have seen earlier, both ONC and SNC require non-zero densities of states at the Fermi energy, NbS$_2$ can be a promising material for the orbital and spin Nernst effects.  

Motivated by this, we first compute the distribution of the orbital and spin Berry curvatures on the \(k_x-k_y\) plane. The distributions are shown in Figs. \ref{fig5}c and d. Similarly to the MoS$_2$ case, predominant contributions to the orbital Berry curvature come from the region surrounding the valley points. We note, however, that in contrast to MoS$_2$, the valley points and their immediate vicinity do not contribute, as those states are unoccupied in NbS$_2$. The predominant contribution to the spin Berry curvature also appears to be from the region surrounding the valley points. We find that the magnitude of the spin Berry curvature is higher for NbS$_2$ compared to that in MoS$_2$, due to the incomplete cancellation of contributions from the oppositely spin-polarized bands. We note that the magnitude of the spin Berry curvature is still smaller compared to the orbital Berry curvature magnitude in NbS$_2$, similar to the MoS$_2$ case. 

Following the orbital and spin Berry curvature distributions, we compute the OHC, ONC, and their spin counterparts, viz., SHC and SNC. Our computed results are shown in Figs. \ref{fig4}b and c.  As expected, we find a non-zero OHC and ONC even in the undoped NbS$_2$ in contrast to MoS$_2$. We also notice a sign change in the OHC and ONC at around 2.43 eV, which is attributed to the sign reversal of the orbital Berry curvature distribution at those corresponding energies (see Appendix \ref{app2}). 

\begin{figure}[t]
    \centering
     \includegraphics[width=\columnwidth]{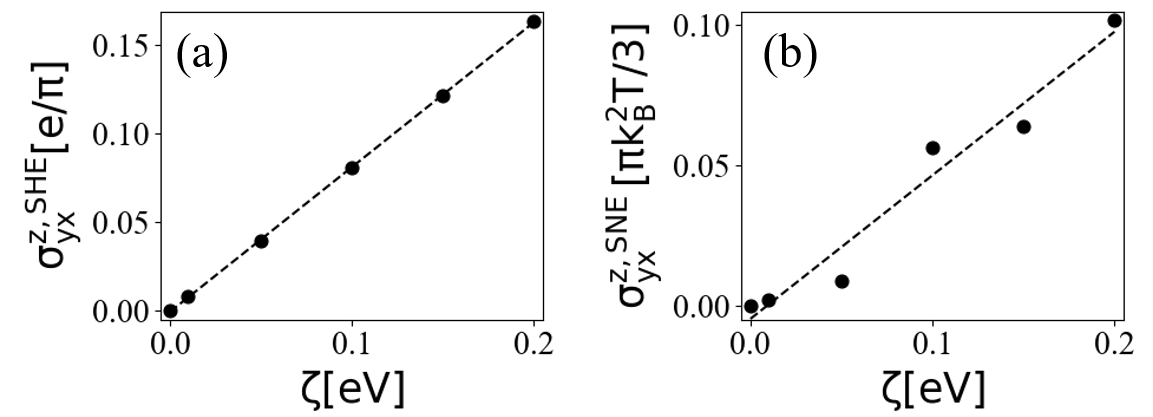}
    \caption{
        Variation of (a) SHC
        in units of \(e/\pi\) and (b) SNC
   in units of \(\pi k_B^2 T/3\) as a function of the spin–orbit coupling constant \(\zeta\) in metallic TMDC NbS\(_2\). The black dashed lines in (a) and (b) represent the linear fitting of the computed SHC and SNC data points.
    }
    \label{fig6}
\end{figure}

Similarly to the OHC and ONC, as shown in Figs. \ref{fig4}b and c, both SHC and SNC are also present in NbS$_2$ even in the absence of doping. We find that generally the orbital-driven effects are higher in magnitude compared to the spin-driven effects. This can be understood from the dependence of both SHC and SNC on the SOC strength, which limits their magnitudes owing to the relatively weak coupling strength of \(\zeta\). As seen from Fig.~\ref{fig6}, both effects are proportional to the SOC constant \(\zeta\) with vanishing magnitude for \(\zeta=0\), consistent with our valley model analysis and also highlighting the importance of the SOC in driving the SHE and SNE.\\

\begin{figure}[t]
    \centering
\includegraphics[width=\columnwidth]{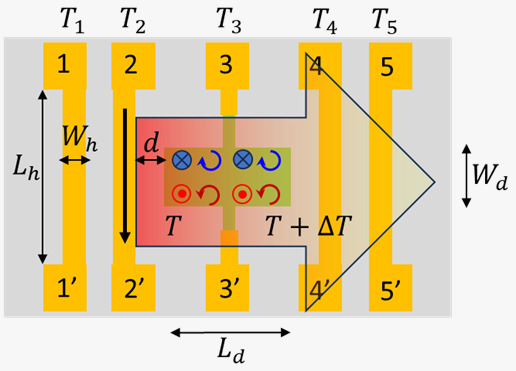}
    \caption{
       The proposed device structure for the detection of SNE and ONE. The monolayer flakes are exfoliated and patterned into Hall-bar shapes. When an electric current is passed through one of the heater lines (11$^\prime$, 22$^\prime$, and so on), which are lithographically fabricated close to the device, an in-plane thermal gradient of the order of 1–10 K$/\mu$m can be created in either direction. The local temperature is monitored by the resistance of all the heater lines and the device. Local magnetic moments arising from spin and orbital contributions accumulate at the edges, shown with closed and open circles, respectively, due to SNE and ONE. These magnetic moments are measured using the MOKE technique.}
    \label{fig7}
\end{figure}

\section{Proposal for Experiments}\label{sec5}

After theoretically demonstrating the presence of the ONE and SNE in monolayer TMDCs, we now discuss possible techniques for probing these effects. Several techniques, including measurements of spin–orbit torques \cite{experiment_1}, spin-Hall magnetoresistance \cite{experiment_2,experiment_3,experiment_4,experiment_5,experiment_6}, and optical detection \cite{experiment_7,experiment_8}, among others \cite{experiment_9,experiment_10,experiment_11}, have been proposed in the literature for detecting orbital currents. As we find from our calculations, SNE and ONE in monolayers of TMDCs generate out-of-plane spin and orbital moments, respectively. These moments flow and eventually accumulate near the device edges along the $y$-axis, transverse to the applied thermal gradient along $\hat{x}$. Here, we discuss the possible detection of the accumulated thermally generated moments using the magneto-optic Kerr effect (MOKE) \cite{experiment_7,experiment_8}. The corresponding device geometry is illustrated in Fig. \ref{fig7}. 

In polar MOKE, a linearly polarized laser beam (typical size 1 $\mu$m) is focused and scanned across the sample surface. Any out-of-plane magnetization causes the polarization axis of the reflected light to rotate by a small angle, a signal known as Kerr rotation. The SNE and ONE generate out-of-plane spin and orbital moments of opposite polarity at the two edges of the channel, which would manifest as Kerr rotations of opposite sign when the beam is scanned from one edge to the other. To confirm the existence of the effect, we can further reverse the direction of the applied thermal gradient. This accordingly flips the polarity of the accumulated moments and, hence, the resulting MOKE signal. Crucially, these measurements can be performed at room temperature in the absence of an external magnetic field. An additional control, however, can be obtained by applying an in-plane magnetic field during detection. This would, in turn, induce Larmor precession of the moments, partially quenching the out-of-plane component and thereby reducing the measured MOKE intensity. Therefore, the in-plane field dependence of the MOKE signal can confirm the existence of SNE and ONE.

Interestingly, for monolayer TMDCs, we find that the polarity of the generated nonequilibrium angular magnetic moments in response to an applied temperature gradient is opposite in sign. Consequently, this leads to a sign reversal in the MOKE signal, allowing us to distinguish between the ONE and the SNE. Notably, a similar systematic analysis of sign changes in measured spin–orbit torques has also been used to disentangle the OHE from the SHE \cite{experiment_12,experiment_13}.

We note that the applied thermal gradient can also generate an electric field due to the Seebeck effect, which, in turn, may lead to electrically induced spin and orbital moment accumulation due to the SHE and OHE, respectively. To separate the contributions of thermally driven spin and orbital currents from their electrical counterparts, a similar experiment can be performed by applying an electric current through the 2D material rather than a thermal gradient. Electrically induced spin and orbital moment accumulation due to the SHE and OHE can also be detected using the same MOKE technique. The applied electric current generates a longitudinal electric field, and we can quantify the normalized Kerr rotation, $\kappa_{J_c}$, per unit electric field. By comparing $\kappa_{J_c}$ with the normalized Kerr rotation, $\kappa_{\nabla T}$, per unit electric field, due to an applied thermal gradient via the Seebeck effect, we can extract the contribution arising solely from the Nernst effect.\\

\section{Summary and Outlook}\label{sec6}

To summarize, our work theoretically demonstrates the presence of orbital and spin currents in monolayer TMDCs in response to an applied temperature gradient and also proposes an experimental setup for their possible detection. Our key findings are based on analytical results from a valley model and more detailed tight-binding calculations, which together allow us to capture the essential physics while also accounting for material-specific features.

Our work highlights the important role of the intrinsic momentum-space orbital moment, Fermi-energy states, and the strength of spin–orbit coupling in governing the magnitude of the ONE and SNE. Such insights are useful for designing materials with desired properties. For example, as revealed by our work, metallicity is one of the crucial criteria for the existence of both ONE and SNE. Consequently, gate-voltage control and electronic doping provide effective knobs to tune both effects, which can also be realized experimentally. Furthermore, the strong SOC dependence of the SNE suggests a route to achieve a larger SNE by using materials with heavy elements. Consequently, monolayers of TaS$_2$ and NbSe$_2$ are expected to exhibit larger SNC than NbS$_2$.

Our work extends the possibility of realizing ONE and SNE to inversion-symmetry-breaking systems by proposing monolayer TMDCs as an excellent material family for hosting these effects. While both metallic and insulating TMDCs have previously been proposed to exhibit OHE, the present work highlights metallic TMDCs as intrinsic hosts for ONE, while insulating TMDCs can also exhibit this effect through extrinsic doping.

Being thermally driven, both ONE and SNE have promising applications in energy harvesting. Fundamentally, these thermally driven effects are also more sensitive to the position of the Fermi level than their electrically driven counterparts, as evident from our analytical results, suggesting their possible application in identifying intricate features in the Fermi surface. This ability to generate orbital and spin currents from heat could open pathways toward new low-power spin–orbitronic and caloritronic devices, where waste heat is recycled into useful information carriers, motivating future studies along these lines.

\section*{Acknowledgements} 
SS and SB thank National Supercomputing Mission for providing computing resources
of ‘PARAM Porul’ at NIT Trichy, and ‘PARAM Rudra’ at IIT Bombay, implemented by C-DAC and supported by the Ministry
of Electronics and Information Technology (MeitY) and Department of Science, and
Technology, Government of India. SB gratefully acknowledges financial support from the IRCC Seed Grant (Project Code: RD/0523-IRCCSH0-018), the INSPIRE Research Grant (Grant No.- DST/INSPIRE/IFF/BATCH-20/2024-25/IFA 23-PH 299), the ANRF PMECRG Grant (Grant No.- ANRF/ECRG/2024/001433/PMS), and the ANRF ARG Grant (Grant No. ARNF/ARG/2025/007161/PS). AB acknowledges the financial support from IITK seed grant (Project Code: 2023578) and the ANRF PMECRG Grant (Project Code: 2025150).\\ 

\appendix

\section{$\Gamma$ Point Analysis}\label{app1}

Here, we discuss the low-energy model around the $\Gamma$ point to understand the large OHC and ONC in the conduction bands of MoS$_2$ and NbS$_2$. As discussed in the main text, both OHC and ONC are governed by the orbital Berry curvature distribution in the momentum space. We, therefore, compute and analyze the orbital Berry curvature for the $\Gamma$ point model Hamiltonian.

To construct the low energy model around the $\Gamma$ point, we first note that in the vicinity of the $\Gamma$ point the valence band is of predominantly \(|d_{3z^2-r^2}\rangle\) orbital character, while the conduction bands have the orbital characters of \(|d_{xz}\rangle\) and \(|d_{yz}\rangle\), respectively.
We, therefore, construct a low-energy model Hamiltonian near the \(\Gamma\) point in the basis set of \( (|d_{xz}\rangle, |d_{3z^2-r^2}\rangle, |d_{yz}\rangle) \), and it is given by 

\begin{widetext}

\[
\mathcal{H}_{\Gamma} (\vec q) = 
\begin{pmatrix}
\Delta +q_x^2a^2t_1+q_y^2a^2t_2& 0 & it_{3}q_xa +q_xq_ya^2t_4\\
0 & \Delta'+q^2a^2t_5 & 0 \\
-it_{3}q_xa +q_xq_ya^2t_4 & 0 & \Delta + q_x^2a^2t_2+q_y^2a^2t_1 \\
\end{pmatrix} \label{Gamma}
\]

\end{widetext}

Here, the parameters, $\Delta, \Delta', $ and $t_i, i=1-5$ are determined by the onsite energies, and the electronic hoppings between different orbitals up to the third nearest neighbor.

\begin{figure}[t]
    \centering
    \includegraphics[width=\columnwidth]{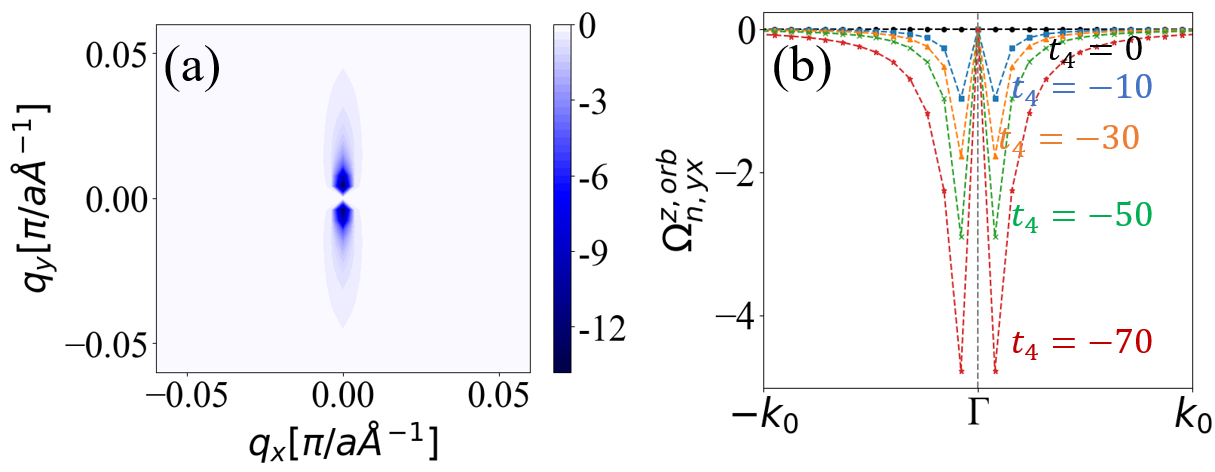}
    \caption{
        Momentum space distribution of the orbital Berry curvature (in units of \(10^4\)\,\AA$^2$), computed for the $\Gamma$ point model, (a) on the $q_x-q_y$ plane, and (b) along $-k_0 \rightarrow \Gamma \rightarrow k_0$ path for different values of \(t_4\).
The values of \(t_4\) in (b) are given in units of meV, and \(k_0 \equiv (0.1\pi/a, 0.1\pi/a) \)
    }
    \label{fig8}
\end{figure}

By diagonalizing the Hamiltonian $\mathcal{H}_{\Gamma}$, we obtain the energy eigenvalues and eigenvectors, and further using them, with the help of Eq. (\ref{obc}), we compute the orbital Berry curvature for the second band in the reciprocal space which constitutes the lowest conduction band in the vicinity of the $\Gamma$ point. The results of our calculations are shown in Fig. \ref{fig8}. As seen from Fig. \ref{fig8}a, the orbital Berry curvature has a large value around the $\Gamma$ point and has the same sign (which is negative in this case) throughout the BZ.

Consequently, summing over these contributions gives us a non-zero value for the OHC, the sign of which is positive due to the additional negative sign in the expression of OHC in Eq. (\ref{ohe_conductivity}). We note that this is consistent with our full BZ results, shown in Figs. \ref{fig3} and \ref{fig4}, where we find that the OHC has a large positive value for both MoS$_2$ and NbS$_2$ in the energy regime of interest.
We further note that the parameter $t_4$, dictated by the inter-orbital $d_{xz}$-$d_{yz}$ hopping parameters, plays a crucial role in determining the orbital Berry curvature. In particular, as seen from Fig. \ref{fig8}b, the magnitude of the orbital Berry curvature increases with increasing value of $t_4$, and also it vanishes in the absence of $t_4$, emphasizing the role of the inter-orbital hopping parameters.

\section{Sign reversal of OHC and ONC}\label{app2}

\begin{figure}[ht]
    \centering
\includegraphics[width=\columnwidth]{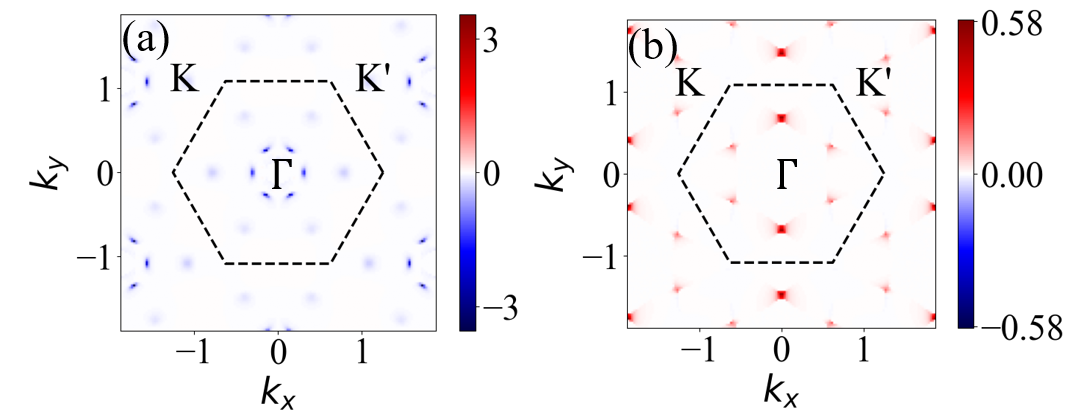}
    \caption{ Plot of orbital Berry curvature for NbS$_2$ in units of $10^3$ \AA$^2$ in the $k_x-k_y$ plane for
        \textbf{(a)} $E_F = 2.43$ eV and for
        \textbf{(b)} $E_F = 3.48$ eV respectively.
    }
    \label{fig9}
\end{figure}

In section \ref{nbs2}, we discuss the sign reversal of the OHC and ONC. As discussed in the main text, the sign reversal can be understood from the analysis of the corresponding orbital Berry curvature distribution. Figs. \ref{fig9}a and b show the $k$-space distributions of the orbital Berry curvature, corresponding to Fermi energies of 2.43 eV and 3.48 eV. We emphasize that at these two Fermi energies, both OHC and ONC have opposite signs. 
As seen from Figs. \ref{fig9}a and b, our computed orbital Berry curvature corresponding to these two Fermi energies 
is opposite in sign. Since the BZ sum of the orbital Berry curvature dictates the OHC, this explains the sign reversal in OHC at these two energies. We further note that there is an overall sign difference in the orbital Berry curvature distributions and the computed OHC, which is also consistent with Eq. (\ref{ohe_conductivity}).
\bibliography{reference}

@article{large_ohe,
  author  = {Sayantika Bhowal and S. Satpathy},
  title   = {Intrinsic orbital moment and prediction of a large orbital Hall effect in two-dimensional transition metal dichalcogenides},
  journal = {Phys. Rev. B},
  volume  = {101},
  pages   = {121112(R)},
  year    = {2020},
  doi     = {10.1103/PhysRevB.101.121112},
}

@article{Canonico2020,
  title = {Orbital Hall insulating phase in transition metal dichalcogenide monolayers},
  author = {Canonico, Luis M. and Cysne, Tarik P. and Molina-Sanchez, Alejandro and Muniz, R. B. and Rappoport, Tatiana G.},
  journal = {Phys. Rev. B},
  volume = {101},
  issue = {16},
  pages = {161409(R)},
  numpages = {6},
  year = {2020},
  month = {Apr},
  publisher = {American Physical Society},
  doi = {10.1103/PhysRevB.101.161409},
  url = {https://link.aps.org/doi/10.1103/PhysRevB.101.161409}
}

@article{Go2018,
  title = {Intrinsic Spin and Orbital Hall Effects from Orbital Texture},
  author = {Go, Dongwook and Jo, Daegeun and Kim, Changyoung and Lee, Hyun-Woo},
  journal = {Phys. Rev. Lett.},
  volume = {121},
  issue = {8},
  pages = {086602},
  numpages = {6},
  year = {2018},
  month = {Aug},
  publisher = {American Physical Society},
  doi = {10.1103/PhysRevLett.121.086602},
  url = {https://link.aps.org/doi/10.1103/PhysRevLett.121.086602}
}

@article{Culter1969,
  title = {Observation of Anderson Localization in an Electron Gas},
  author = {Cutler, Melvin and Mott, N. F.},
  journal = {Phys. Rev.},
  volume = {181},
  issue = {3},
  pages = {1336--1340},
  numpages = {0},
  year = {1969},
  month = {May},
  publisher = {American Physical Society},
  doi = {10.1103/PhysRev.181.1336},
  url = {https://link.aps.org/doi/10.1103/PhysRev.181.1336}
}

@article{Hall1880,
author = {E.H. Hall},
title = {XXXVIII. On the new action of magnetism on a permanent electric current },
journal = {The London, Edinburgh, and Dublin Philosophical Magazine and Journal of Science},
volume = {10},
number = {63},
pages = {301--328},
year = {1880},
publisher = {Taylor \& Francis},
doi = {10.1080/14786448008626936},
URL = {     
        https://doi.org/10.1080/14786448008626936
},
eprint = { 
        https://doi.org/10.1080/14786448008626936
}
}

@article{Hall1881,
author = {E.H. Hall},
title = {XVIII. On the “Rotational Coefficient” in nickel and cobalt },
journal = {The London, Edinburgh, and Dublin Philosophical Magazine and Journal of Science},
volume = {12},
number = {74},
pages = {157--172},
year = {1881},
publisher = {Taylor \& Francis},
doi = {10.1080/14786448108627086},
URL = {     
        https://doi.org/10.1080/14786448108627086
},
eprint = { 
        https://doi.org/10.1080/14786448108627086
}
}

@article{gapped_graphene,
  title = {Orbital Hall effect as an alternative to valley Hall effect in gapped graphene},
  author = {Bhowal, Sayantika and Vignale, Giovanni},
  journal = {Phys. Rev. B},
  volume = {103},
  issue = {19},
  pages = {195309},
  numpages = {8},
  year = {2021},
  month = {May},
  publisher = {American Physical Society},
  doi = {10.1103/PhysRevB.103.195309},
  url = {https://link.aps.org/doi/10.1103/PhysRevB.103.195309}
}

@article{Nernst1886,
author = {A. v. Ettingshausen and W. Nernst
},
title = {Ueber das Auftreten electromotorischer Kräfte in Metallplatten, welche von einem Wärmestrome durchflossen werden und sich im magnetischen Felde befinden },
journal = {Annalen der Physik},
volume = {265},
number = {10},
pages = {343--347},
year = {1886},
doi = {10.1002/andp.18862651010},
URL = {     
        https://doi.org/10.1002/andp.18862651010
},
}

@article{quantum_devices1,
  title = {Nondissipative Spin Hall Effect via Quantized Edge Transport},
  author = {Sheng, L. and Sheng, D. N. and Ting, C. S. and Haldane, F. D. M.},
  journal = {Phys. Rev. Lett.},
  volume = {95},
  issue = {13},
  pages = {136602},
  numpages = {4},
  year = {2005},
  month = {Sep},
  publisher = {American Physical Society},
  doi = {10.1103/PhysRevLett.95.136602},
  url = {https://link.aps.org/doi/10.1103/PhysRevLett.95.136602}
}

@article{quantum_devices2,
author = {Manuel Meyer  and Jonas Baumbach  and Sergey Krishtopenko  and Adriana Wolf  and Monika Emmerling  and Sebastian Schmid  and Martin Kamp  and Benoit Jouault  and Jean-Baptiste Rodriguez  and Eric Tournie  and Tobias Müller  and Ronny Thomale  and Gerald Bastard  and Frederic Teppe  and Fabian Hartmann  and Sven Höfling },
title = {Quantum spin Hall effect in III-V semiconductors at elevated temperatures: Advancing topological electronics},
journal = {Science Advances},
volume = {11},
number = {43},
pages = {eadz2408},
year = {2025},
doi = {10.1126/sciadv.adz2408},
URL = {https://www.science.org/doi/abs/10.1126/sciadv.adz2408},
}

@article{quantum_devices3,
  title = {Two-dimensional orbital Hall insulators},
  author = {Canonico, Luis M. and Cysne, Tarik P. and Rappoport, Tatiana G. and Muniz, R. B.},
  journal = {Phys. Rev. B},
  volume = {101},
  issue = {7},
  pages = {075429},
  numpages = {5},
  year = {2020},
  month = {Feb},
  publisher = {American Physical Society},
  doi = {10.1103/PhysRevB.101.075429},
  url = {https://link.aps.org/doi/10.1103/PhysRevB.101.075429}
}

@article{spin_hall_nernst,
title = {Anisotropic spin Hall and spin Nernst effects in bismuth semimetal},
journal = {Journal of Magnetism and Magnetic Materials},
volume = {563},
pages = {169949},
year = {2022},
issn = {0304-8853},
doi = {https://doi.org/10.1016/j.jmmm.2022.169949},
url = {https://www.sciencedirect.com/science/article/pii/S0304885322008344},
author = {Guang-Yu Guo},

}

@article{
spin_hall_new,
author = {Y. K. Kato  and R. C. Myers  and A. C. Gossard  and D. D. Awschalom },
title = {Observation of the Spin Hall Effect in Semiconductors},
journal = {Science},
volume = {306},
number = {5703},
pages = {1910-1913},
year = {2004},
doi = {10.1126/science.1105514},
URL = {https://www.science.org/doi/abs/10.1126/science.1105514}
}

@article{heavy_spin_hall2,
  title = {Indication of intrinsic spin Hall effect in $4d$ and $5d$ transition metals},
  author = {Morota, M. and Niimi, Y. and Ohnishi, K. and Wei, D. H. and Tanaka, T. and Kontani, H. and Kimura, T. and Otani, Y.},
  journal = {Phys. Rev. B},
  volume = {83},
  issue = {17},
  pages = {174405},
  numpages = {5},
  year = {2011},
  month = {May},
  publisher = {American Physical Society},
  doi = {10.1103/PhysRevB.83.174405},
  url = {https://link.aps.org/doi/10.1103/PhysRevB.83.174405}
}

@inbook{Kittel2004,
  author    = {Charles Kittel},
  title     = {Surface and Interface Physics},
  booktitle = {Introduction to Solid State Physics},
  publisher = {Wiley},
  address   = {New York},
  year      = {2004},
  chapter   = {17},
  pages     = {487--514}
}

@article{orbital_hall_experiment1,
  title = {Orbital Hall physics in two-dimensional Dirac materials},
  author = {Pezo, Armando and Garc\'{\i}a Ovalle, Diego and Manchon, Aur\'elien},
  journal = {Phys. Rev. B},
  volume = {108},
  issue = {7},
  pages = {075427},
  numpages = {10},
  year = {2023},
  month = {Aug},
  publisher = {American Physical Society},
  doi = {10.1103/PhysRevB.108.075427},
  url = {https://link.aps.org/doi/10.1103/PhysRevB.108.075427}
}

@article{orbital_hall_experiment2,
  title = {Observation of long-range orbital transport and
  giant orbital torque},
  author = {Hiroki Hayashi and Daegeun Jo and Dongwook Go and Tenghua Gao and Satoshi Haku and Yuriy Mokrousov and
 Hyun-Woo Lee and Kazuya Ando },
  journal = {},
  volume = {},
  issue = {},
  pages = {},
  numpages = {},
  year = {2023},
  month = {},
  publisher = {},
  doi = {10.1038/s42005-023-01139-7},
  url = {https://doi.org/10.1038/s42005-023-01139-7}
}

@article{ohe,
  author  = {Sayantika Bhowal and S. Satpathy},
  title   = {Intrinsic orbital and spin Hall effects in monolayer transition metal dichalcogenides},
  journal = {Phys. Rev. B},
  volume  = {102},
  pages   = {035409},
  year    = {2020},
  doi     = {10.1103/PhysRevB.102.035409},
}

@article{Tarik2021,
  title = {Disentangling Orbital and Valley Hall Effects in Bilayers of Transition Metal Dichalcogenides},
  author = {Cysne, Tarik P. and Costa, Marcio and Canonico, Luis M. and Nardelli, M. Buongiorno and Muniz, R. B. and Rappoport, Tatiana G.},
  journal = {Phys. Rev. Lett.},
  volume = {126},
  issue = {5},
  pages = {056601},
  numpages = {7},
  year = {2021},
  month = {Feb},
  publisher = {American Physical Society},
  doi = {10.1103/PhysRevLett.126.056601},
  url = {https://link.aps.org/doi/10.1103/PhysRevLett.126.056601}
}

@article{Cysne2021,
  title = {Orbital magnetoelectric effect in zigzag nanoribbons of $p$-band systems},
  author = {Cysne, Tarik P. and Guimar\~aes, Filipe S. M. and Canonico, Luis M. and Rappoport, Tatiana G. and Muniz, R. B.},
  journal = {Phys. Rev. B},
  volume = {104},
  issue = {16},
  pages = {165403},
  numpages = {8},
  year = {2021},
  month = {Oct},
  publisher = {American Physical Society},
  doi = {10.1103/PhysRevB.104.165403},
  url = {https://link.aps.org/doi/10.1103/PhysRevB.104.165403}
}

@article{Sala2022,
  title = {Giant orbital Hall effect and orbital-to-spin conversion in $3d$, $5d$, and $4f$ metallic heterostructures},
  author = {Sala, Giacomo and Gambardella, Pietro},
  journal = {Phys. Rev. Res.},
  volume = {4},
  issue = {3},
  pages = {033037},
  numpages = {14},
  year = {2022},
  month = {Jul},
  publisher = {American Physical Society},
  doi = {10.1103/PhysRevResearch.4.033037},
  url = {https://link.aps.org/doi/10.1103/PhysRevResearch.4.033037}
}

@article{Sala2023,
  title = {Orbital Hanle Magnetoresistance in a $3d$ Transition Metal},
  author = {Sala, Giacomo and Wang, Hanchen and Legrand, William and Gambardella, Pietro},
  journal = {Phys. Rev. Lett.},
  volume = {131},
  issue = {15},
  pages = {156703},
  numpages = {6},
  year = {2023},
  month = {Oct},
  publisher = {American Physical Society},
  doi = {10.1103/PhysRevLett.131.156703},
  url = {https://link.aps.org/doi/10.1103/PhysRevLett.131.156703}
}

@article{Bhowal2020,
  title = {Orbital gyrotropic magnetoelectric effect and its strain engineering in monolayer $\mathrm{Nb}{X}_{2}$},
  author = {Bhowal, Sayantika and Satpathy, S.},
  journal = {Phys. Rev. B},
  volume = {102},
  issue = {20},
  pages = {201403(R)},
  numpages = {6},
  year = {2020},
  month = {Nov},
  publisher = {American Physical Society},
  doi = {10.1103/PhysRevB.102.201403},
  url = {https://link.aps.org/doi/10.1103/PhysRevB.102.201403}
}

@article{Salemi2019,
	author = {Salemi, Leandro and Berritta, Marco and Nandy, Ashis K. and Oppeneer, Peter M.},
	date = {2019/11/26},
	doi = {10.1038/s41467-019-13367-z},
	id = {Salemi2019},
	isbn = {2041-1723},
	journal = {Nature Communications},
	number = {1},
	pages = {5381},
	title = {Orbitally dominated Rashba-Edelstein effect in noncentrosymmetric antiferromagnets},
	url = {https://doi.org/10.1038/s41467-019-13367-z},
	volume = {10},
	year = {2019}}

@misc{Sun2024,
      title={Orbital Magnetic Moment Dynamics and Hanle Magnetoresistance in Multilayered 2D Materials}, 
      author={Hao Sun and Giovanni Vignale},
      year={2024},
      eprint={2408.02887},
      archivePrefix={arXiv},
      primaryClass={cond-mat.mes-hall},
      url={https://arxiv.org/abs/2408.02887}, 
}

@article{Cysne2025,
	author = {Cysne, Tarik P. and Canonico, Luis M. and Costa, Marcio and Muniz, R. B. and Rappoport, Tatiana G.},
	date = {2025/10/01},
	doi = {10.1038/s44306-025-00103-1},
	id = {Cysne2025},
	isbn = {2948-2119},
	journal = {npj Spintronics},
	number = {1},
	pages = {39},
	title = {Orbitronics in two-dimensional materials},
	url = {https://doi.org/10.1038/s44306-025-00103-1},
	volume = {3},
	year = {2025}}

@article{Tarik2022,
  title = {Orbital Hall effect in bilayer transition metal dichalcogenides: From the intra-atomic approximation to the Bloch states orbital magnetic moment approach},
  author = {Cysne, Tarik P. and Bhowal, Sayantika and Vignale, Giovanni and Rappoport, Tatiana G.},
  journal = {Phys. Rev. B},
  volume = {105},
  issue = {19},
  pages = {195421},
  numpages = {15},
  year = {2022},
  month = {May},
  publisher = {American Physical Society},
  doi = {10.1103/PhysRevB.105.195421},
  url = {https://link.aps.org/doi/10.1103/PhysRevB.105.195421}
}

@misc{Dutta2026,
      title={Interplay of Valley, Orbital, Spin, and Layer Degrees of Freedom in Ta$_2$CS$_2$ MXene}, 
      author={Kunal Dutta and Anupam Mondal and Sayantika Bhowal and Subhradip Ghosh and Indra Dasgupta},
      year={2026},
      eprint={2605.01271},
      archivePrefix={arXiv},
      primaryClass={cond-mat.str-el},
      url={https://arxiv.org/abs/2605.01271}, 
}

@article{new_orbitalhall2,
  title = {Imaging the valley and orbital Hall effect in monolayer ${\mathrm{MoS}}_{2}$},
  author = {Xue, Fei and Amin, Vivek and Haney, Paul M.},
  journal = {Phys. Rev. B},
  volume = {102},
  issue = {16},
  pages = {161103},
  numpages = {5},
  year = {2020},
  month = {Oct},
  publisher = {American Physical Society},
  doi = {10.1103/PhysRevB.102.161103},
  url = {https://link.aps.org/doi/10.1103/PhysRevB.102.161103}
}

@article{new_orbitalhall3,
  title = {Orbitronics: The Intrinsic Orbital Current in $p$-Doped Silicon},
  author = {Bernevig, B. Andrei and Hughes, Taylor L. and Zhang, Shou-Cheng},
  journal = {Phys. Rev. Lett.},
  volume = {95},
  issue = {6},
  pages = {066601},
  numpages = {4},
  year = {2005},
  month = {Aug},
  publisher = {American Physical Society},
  doi = {10.1103/PhysRevLett.95.066601},
  url = {https://link.aps.org/doi/10.1103/PhysRevLett.95.066601}
}

@article{one,
  author  = {Leandro Salemi and Peter M. Oppeneer},
  title   = {First-principles theory of intrinsic spin and orbital Hall and Nernst effects in metallic monoatomic crystals},
  journal = {Phys. Rev. Materials},
  volume  = {6},
  pages   = {095001},
  year    = {2022},
  doi     = {10.1103/PhysRevMaterials.6.095001},
}

@article{gigantic,
  author  = {Daegeun Jo and Dongwook Go and Hyun-Woo Lee},
  title   = {Gigantic intrinsic orbital Hall effects in weakly spin-orbit coupled metals},
  journal = {Phys. Rev. B},
  volume  = {98},
  pages   = {214405},
  year    = {2018},
  doi     = {10.1103/PhysRevB.98.214405},
}

@article{berry1,
  author  = {S. Pancharatnam},
  title   = {Generalized Theory of Interference, and Its Applications},
  journal = {Proceedings of the Indian Academy of Sciences, Section A},
  volume  = {44},
  pages   = {},
  year    = {1956},
  doi     = {10.1007/BF03046095},
}

@article{berry2,
  author  = {C. Alden Mead},
  title   = {The geometric phase in molecular systems},
  journal = {Rev. Mod. Phys.},
  volume  = {64},
  pages   = {51},
  year    = {1992},
  doi     = {10.1103/RevModPhys.64.51},
}

@article{berry3,
  author  = {Di Xiao and Ming-Che Chang and Qian Niu},
  title   = {Berry phase effects on electronic properties},
  journal = {Rev. Mod. Phys.},
  volume  = {82},
  pages   = {1959},
  year    = {2010},
  doi     = {10.1103/RevModPhys.82.1959},
}

@article{berry4,
  author  = {D. J. Thouless},
  title   = {Localisation and the two-dimensional Hall effect},
  journal = {J. Phys. C: Solid State Phys.},
  volume  = {14},
  pages   = {},
  year    = {1981},
  doi     = {10.1088/0022-3719/14/23/022},
}

@article{berry5,
  author  = {T. Jungwirth and Qian Niu and A. H. MacDonald},
  title   = {Anomalous Hall Effect in Ferromagnetic Semiconductors},
  journal = {Phys. Rev. Lett.},
  volume  = {88},
  pages   = {207208},
  year    = {2002},
  doi     = {10.1103/PhysRevLett.88.207208},
}

@article{berry6,
  author  = {T. Thonhauser and Davide Ceresoli and David Vanderbilt and R. Resta},
  title   = {Orbital Magnetization in Periodic Insulators},
  journal = {Phys. Rev. Lett.},
  volume  = {95},
  pages   = {137205},
  year    = {2005},
  doi     = {10.1103/PhysRevLett.95.137205},
}

@article{berry7,
  author  = {Konstantin Y. Bliokh and Avi Niv and Vladimir Kleiner and Erez Hasman},
  title   = {Geometrodynamics of spinning light},
  journal = {Nature Photonics},
  volume  = {2},
  pages   = {},
  year    = {2008},
  doi     = {10.1038/nphoton.2008.229},
}

@article{orbitronics_1,
  title = {Orbitronics: The Intrinsic Orbital Current in $p$-Doped Silicon},
  author = {Bernevig, B. Andrei and Hughes, Taylor L. and Zhang, Shou-Cheng},
  journal = {Phys. Rev. Lett.},
  volume = {95},
  issue = {6},
  pages = {066601},
  numpages = {4},
  year = {2005},
  month = {Aug},
  publisher = {American Physical Society},
  doi = {10.1103/PhysRevLett.95.066601},
  url = {https://link.aps.org/doi/10.1103/PhysRevLett.95.066601}
}

@article{orbitronics_2,
  title = {Optically Controlled Orbitronics on a Triangular Lattice},
  author = {Phong, V\~o Tiến and Addison, Zachariah and Ahn, Seongjin and Min, Hongki and Agarwal, Ritesh and Mele, E. J.},
  journal = {Phys. Rev. Lett.},
  volume = {123},
  issue = {23},
  pages = {236403},
  numpages = {6},
  year = {2019},
  month = {Dec},
  publisher = {American Physical Society},
  doi = {10.1103/PhysRevLett.123.236403},
  url = {https://link.aps.org/doi/10.1103/PhysRevLett.123.236403}
}

@article{quantum_hall1,
  author    = {von Klitzing, Klaus and Chakraborty, Tapash and Kim, Philip and others},
  title     = {40 years of the quantum Hall effect},
  journal   = {Nature Reviews Physics},
  volume    = {2},
  number    = {7},
  pages     = {397--401},
  year      = {2020},
  doi       = {10.1038/s42254-020-0209-1},
  url       = {https://doi.org/10.1038/s42254-020-0209-1}
}

@article{spintronics1,
  title = {Spintronics: Fundamentals and applications},
  author = {\ifmmode \check{Z}\else \v{Z}\fi{}uti\ifmmode \acute{c}\else \'{c}\fi{}, Igor and Fabian, Jaroslav and Das Sarma, S.},
  journal = {Rev. Mod. Phys.},
  volume = {76},
  issue = {2},
  pages = {323--410},
  numpages = {0},
  year = {2004},
  month = {Apr},
  publisher = {American Physical Society},
  doi = {10.1103/RevModPhys.76.323},
  url = {https://link.aps.org/doi/10.1103/RevModPhys.76.323}
}

@article{spintronics2,
  title = {Nobel Lecture: Origin, development, and future of spintronics},
  author = {Fert, Albert},
  journal = {Rev. Mod. Phys.},
  volume = {80},
  issue = {4},
  pages = {1517--1530},
  numpages = {0},
  year = {2008},
  month = {Dec},
  publisher = {American Physical Society},
  doi = {10.1103/RevModPhys.80.1517},
  url = {https://link.aps.org/doi/10.1103/RevModPhys.80.1517}
}

@article{orbitronics_3,
  author    = {Fukami, Shunsuke and Lee, K.-J. and Kl{\"a}ui, Mathias},
  title     = {Challenges and opportunities in orbitronics},
  journal   = {Nature Physics},
  year      = {2025},
  doi       = {10.1038/s41567-025-03143-w},
  url       = {https://doi.org/10.1038/s41567-025-03143-w}
}

@article{quantum_orbital_magnetization,
  title = {Quantum Theory of Orbital Magnetization and Its Generalization to Interacting Systems},
  author = {Shi, Junren and Vignale, G. and Xiao, Di and Niu, Qian},
  journal = {Phys. Rev. Lett.},
  volume = {99},
  issue = {19},
  pages = {197202},
  numpages = {4},
  year = {2007},
  month = {Nov},
  publisher = {American Physical Society},
  doi = {10.1103/PhysRevLett.99.197202},
  url = {https://link.aps.org/doi/10.1103/PhysRevLett.99.197202}
}

@article{pure_spin_current,
  title = {Manipulation of pure spin current in ferromagnetic metals independent of magnetization},
  author = {Tian, Dai and Li, Yufan and Qu, D. and Huang, S. Y. and Jin, Xiaofeng and Chien, C. L.},
  journal = {Phys. Rev. B},
  volume = {94},
  issue = {2},
  pages = {020403(R)},
  numpages = {4},
  year = {2016},
  month = {Jul},
  publisher = {American Physical Society},
  doi = {10.1103/PhysRevB.94.020403},
  url = {https://link.aps.org/doi/10.1103/PhysRevB.94.020403}
}

@article{intrinsic_spin_hall,
  title = {Universal Intrinsic Spin Hall Effect},
  author = {Sinova, Jairo and Culcer, Dimitrie and Niu, Q. and Sinitsyn, N. A. and Jungwirth, T. and MacDonald, A. H.},
  journal = {Phys. Rev. Lett.},
  volume = {92},
  issue = {12},
  pages = {126603},
  numpages = {4},
  year = {2004},
  month = {Mar},
  publisher = {American Physical Society},
  doi = {10.1103/PhysRevLett.92.126603},
  url = {https://link.aps.org/doi/10.1103/PhysRevLett.92.126603}
}

@article{giant_spin_hall,
  author    = {Derunova, Elizaveta and Sun, Yin and Felser, Claudia and Parkin, Stuart S. P. and Yan, Binghai},
  title     = {Giant intrinsic spin Hall effect in W$_3$Ta and other A15 superconductors},
  journal   = {Science Advances},
  volume    = {5},
  number    = {5},
  pages     = {eaav8575},
  year      = {2019},
  doi       = {10.1126/sciadv.aav8575},
  url       = {https://doi.org/10.1126/sciadv.aav8575}
}

@article{strong_intrinsic_spin_hall,
  title = {Strong Intrinsic Spin Hall Effect in the TaAs Family of Weyl Semimetals},
  author = {Sun, Yan and Zhang, Yang and Felser, Claudia and Yan, Binghai},
  journal = {Phys. Rev. Lett.},
  volume = {117},
  issue = {14},
  pages = {146403},
  numpages = {5},
  year = {2016},
  month = {Sep},
  publisher = {American Physical Society},
  doi = {10.1103/PhysRevLett.117.146403},
  url = {https://link.aps.org/doi/10.1103/PhysRevLett.117.146403}
}

@article{greater_orbital_hall,
  title = {Dominance of the orbital Hall effect over spin in transition metal heterostructures},
  author = {Costa, J. L. and Santos, E. and Mendes, J. B. S. and Azevedo, A.},
  journal = {Phys. Rev. B},
  volume = {112},
  issue = {5},
  pages = {054443},
  numpages = {9},
  year = {2025},
  month = {Aug},
  publisher = {American Physical Society},
  doi = {10.1103/tbdz-jc6k},
  url = {https://link.aps.org/doi/10.1103/tbdz-jc6k}
}

@article{OrbitalTransport,
  author    = {Yang, Q. and Xiao, J. and Robredo, I. and Vergniory, M. G. and Yan, B. and Felser, C.},
  title     = {Monopole-like orbital-momentum locking and the induced orbital transport in topological chiral semimetals},
  journal   = {Proceedings of the National Academy of Sciences of the United States of America},
  volume    = {120},
  number    = {48},
  pages     = {e2305541120},
  year      = {2023},
  doi       = {10.1073/pnas.2305541120},
  url       = {https://doi.org/10.1073/pnas.2305541120}
}

@article{orbital_transport2,
  title = {Valley-dependent giant orbital moments and transport features in rhombohedral graphene multilayers},
  author = {Mu, Xingchi and Zhou, Jian},
  journal = {Phys. Rev. B},
  volume = {111},
  issue = {16},
  pages = {165102},
  numpages = {10},
  year = {2025},
  month = {Apr},
  publisher = {American Physical Society},
  doi = {10.1103/PhysRevB.111.165102},
  url = {https://link.aps.org/doi/10.1103/PhysRevB.111.165102}
}

@article{orbital_transport_3,
  title = {Negative intrinsic orbital Hall effect in group XIV materials},
  author = {Baek, Insu and Lee, Hyun-Woo},
  journal = {Phys. Rev. B},
  volume = {104},
  issue = {24},
  pages = {245204},
  numpages = {7},
  year = {2021},
  month = {Dec},
  publisher = {American Physical Society},
  doi = {10.1103/PhysRevB.104.245204},
  url = {https://link.aps.org/doi/10.1103/PhysRevB.104.245204}
}

@article{centrosymmetric1,
  title = {Emergence of giant orbital Hall and tunable spin Hall effects in centrosymmetric transition metal dichalcogenides},
  author = {Sahu, Pratik and Bidika, Jatin Kumar and Biswal, Bubunu and Satpathy, S. and Nanda, B. R. K.},
  journal = {Phys. Rev. B},
  volume = {110},
  issue = {5},
  pages = {054403},
  numpages = {8},
  year = {2024},
  month = {Aug},
  publisher = {American Physical Society},
  doi = {10.1103/PhysRevB.110.054403},
  url = {https://link.aps.org/doi/10.1103/PhysRevB.110.054403}
}

@article{normalspinhall,
  author  = {Jairo Sinova and Sergio O. Valenzuela and J. Wunderlich and C. H. Back and T. Jungwirth},
  title   = {Spin Hall effects},
  journal = {Rev. Mod. Phys.},
  volume  = {87},
  pages   = {1213},
  year    = {2015},
  doi     = {10.1103/RevModPhys.87.1213},
}

@article{normalspinnernst,
  author  = {Atsuo Shitade},
  title   = {Spin accumulation in the spin Nernst effect},
  journal = {Phys. Rev. B},
  volume  = {106},
  pages   = {045203},
  year    = {2022},
  doi     = {10.1103/PhysRevB.106.045203},
}

@article{tmdc1,
  author  = {Chumki Nayak and Suvadip Masanta and Sukanya Ghosh and Shubhadip Moulick and Atindra Nath Pal and Indrani Bose and Achintya Singha},
  title   = {Valley polarization and photocurrent generation in transition metal dichalcogenide alloy {MoS$_{2x}$Se$_{2(1-x)}$}},
  journal = {Phys. Rev. B},
  volume  = {109},
  pages   = {115304},
  year    = {2024},
  doi     = {10.1103/PhysRevB.109.115304},
}

@article{tmdc2,
  author  = {Alka Rani and Arpit Verma and Bal Chandra Yadav},
  title   = {Advancements in transition metal dichalcogenides (TMDCs) for self-powered photodetectors: challenges, properties, and functionalization strategies},
  journal = {Materials Advances},
  volume  = {6},
  pages   = {},
  year    = {2024},
  doi     = {10.1039/D3MA01152F},
}

@article{nmto5,
  author  = {O. K. Andersen and T. Saha-Dasgupta},
  title   = {Muffin-tin orbitals of arbitrary order},
  journal = {Physical Review B},
  volume  = {62},
  pages   = {R16219--R16222},
  year    = {2000},
  doi     = {10.1103/PhysRevB.62.R16219},
}

@article{verylargeorbital3,
  author  = {Pratik Sahu and Sayantika Bhowal and S. Satpathy},
  title   = {Effect of the inversion symmetry breaking on the orbital Hall effect: A model study},
  journal = {Physical Review B},
  volume  = {103},
  pages   = {085113},
  year    = {2021},
  doi     = {10.1103/PhysRevB.103.085113},
}

@article{experiment_1,
  title = {Current-induced spin-orbit torques in ferromagnetic and antiferromagnetic systems},
  author = {Manchon, A. and \ifmmode \check{Z}\else \v{Z}\fi{}elezn\'y, J. and Miron, I. M. and Jungwirth, T. and Sinova, J. and Thiaville, A. and Garello, K. and Gambardella, P.},
  journal = {Rev. Mod. Phys.},
  volume = {91},
  issue = {3},
  pages = {035004},
  numpages = {80},
  year = {2019},
  month = {Sep},
  publisher = {American Physical Society},
  doi = {10.1103/RevModPhys.91.035004},
  url = {https://link.aps.org/doi/10.1103/RevModPhys.91.035004}
}

@article{experiment_2,
  author  = {Meyer, S. and Chen, Y. T. and Wimmer, S. and Althammer, M. and Wimmer, T. and Schlitz, R. and Geprägs, S. and Huebl, H. and Ködderitzsch, D. and Ebert, H. and Bauer, G. E. W. and Gross, R. and Goennenwein, S. T. B.},
  title   = {Observation of the spin Nernst effect},
  journal = {Nature Materials},
  year    = {2017},
  volume  = {16},
  number  = {10},
  pages   = {977--981},
  doi     = {10.1038/nmat4964},
  note    = {Epub 2017 Sep 11},
  pmid    = {28892056}
}

@article{experiment_3,
author = {Sheng, Peng and Sakuraba, Yuya and Lau, Yongchang and Takahashi, Saburo and Mitani, Seiji and Hayashi, Masamitsu},
year = {2017},
month = {11},
pages = {e1701503},
title = {The spin Nernst effect in tungsten},
volume = {3},
journal = {Science Advances},
doi = {10.1126/sciadv.1701503}
}

@article{experiment_4,
  author  = {Kim, D. J. and Jeon, C. Y. and Choi, J. G. and Lee, J. W. and Surabhi, S. and Jeong, J. R. and Lee, K. J. and Park, B. G.},
  title   = {Observation of transverse spin Nernst magnetoresistance induced by thermal spin current in ferromagnet/non-magnet bilayers},
  journal = {Nature Communications},
  year    = {2017},
  volume  = {8},
  number  = {1},
  pages   = {1400},
  doi     = {10.1038/s41467-017-01493-5},
  note    = {Erratum in: Nature Communications (2018), 9(1):138, doi:10.1038/s41467-017-02303-8},
  pmid    = {29123123},
  pmcid   = {PMC5680200}
}

@article{experiment_5,
  title = {Origin of transverse voltages generated by thermal gradients and electric fields in ferrimagnetic-insulator/heavy-metal bilayers},
  author = {Bose, Arnab and Jain, Rakshit and Bauer, Jackson J. and Buhrman, Robert A. and Ross, Caroline A. and Ralph, Daniel C.},
  journal = {Phys. Rev. B},
  volume = {105},
  issue = {10},
  pages = {L100408},
  numpages = {6},
  year = {2022},
  month = {Mar},
  publisher = {American Physical Society},
  doi = {10.1103/PhysRevB.105.L100408},
  url = {https://link.aps.org/doi/10.1103/PhysRevB.105.L100408}
}

@article{experiment_6,
  author  = {Jain, R. and Stanley, M. and Bose, A. and Richardella, A. R. and Zhang, X. S. and Pillsbury, T. and Muller, D. A. and Samarth, N. and Ralph, D. C.},
  title   = {Thermally generated spin current in the topological insulator Bi$_2$Se$_3$},
  journal = {Science Advances},
  year    = {2023},
  volume  = {9},
  number  = {50},
  pages   = {eadi4540},
  doi     = {10.1126/sciadv.adi4540},
  note    = {Epub 2023 Dec 13},
  pmid    = {38091392},
  pmcid   = {PMC10848729}
}

@article{experiment_7,
  author  = {Kato, Y. K. and Myers, R. C. and Gossard, A. C. and Awschalom, D. D.},
  title   = {Observation of the spin Hall effect in semiconductors},
  journal = {Science},
  year    = {2004},
  volume  = {306},
  number  = {5703},
  pages   = {1910--1913},
  doi     = {10.1126/science.1105514},
  note    = {Epub 2004 Nov 11},
  pmid    = {15539563}
}

@article{experiment_8,
  author  = {Choi, Y. G. and Jo, D. and Ko, K. H. and Go, D. and Kim, K. H. and Park, H. G. and Kim, C. and Min, B. C. and Choi, G. M. and Lee, H. W.},
  title   = {Observation of the orbital Hall effect in a light metal Ti},
  journal = {Nature},
  year    = {2023},
  volume  = {619},
  number  = {7968},
  pages   = {52--56},
  doi     = {10.1038/s41586-023-06101-9},
  note    = {Epub 2023 Jul 5},
  pmid    = {37407680}
}

@article{experiment_9,
author = {Bose, Arnab and Bhuktare, Swapnil and Singh, Hanuman and Dutta, Sutapa and Achanta, Venu Gopal and Tulapurkar, Ashwin},
year = {2018},
month = {04},
pages = {162401},
title = {Direct detection of spin Nernst effect in platinum},
volume = {112},
journal = {Applied Physics Letters},
doi = {10.1063/1.5021731}
}

@article{experiment_10,
  title = {Control of magnetization dynamics by spin-Nernst torque},
  author = {Bose, Arnab and Shukla, Ambika Shanker and Dutta, Sutapa and Bhuktare, Swapnil and Singh, Hanuman and Tulapurkar, Ashwin A.},
  journal = {Phys. Rev. B},
  volume = {98},
  issue = {18},
  pages = {184412},
  numpages = {8},
  year = {2018},
  month = {Nov},
  publisher = {American Physical Society},
  doi = {10.1103/PhysRevB.98.184412},
  url = {https://link.aps.org/doi/10.1103/PhysRevB.98.184412}
}

@article{experiment_11,
  author  = {Kim, J. M. and Kim, D. J. and Cheon, C. Y. and Moon, K. W. and Kim, C. and Cao Van, P. and Jeong, J. R. and Hwang, C. and Lee, K. J. and Park, B. G.},
  title   = {Observation of Thermal Spin-Orbit Torque in W/CoFeB/MgO Structures},
  journal = {Nano Letters},
  year    = {2020},
  volume  = {20},
  number  = {11},
  pages   = {7803--7810},
  doi     = {10.1021/acs.nanolett.0c01702},
  note    = {Epub 2020 Oct 15},
  pmid    = {33054243}
}

@article{experiment_12,
  author  = {Lee, Dongjoon and Go, Dongwook and Park, Hyeon-Jong and Jeong, Wonmin and Ko, Hye-Won and Yun, Deokhyun and Jo, Daegeun and Lee, Soogil and Go, Gyungchoon and Oh, Jung Hyun and Kim, Kab-Jin and Park, Byong-Guk and Min, Byoung-Chul and Koo, Hyun Cheol and Lee, Hyun-Woo and Lee, OukJae and Lee, Kyung-Jin},
  title   = {Orbital torque in magnetic bilayers},
  journal = {Nature Communications},
  year    = {2021},
  volume  = {12},
  number  = {1},
  pages   = {6710},
  doi     = {10.1038/s41467-021-26650-9}
}

@article{experiment_13,
  title = {Detection of long-range orbital-Hall torques},
  author = {Bose, Arnab and Kammerbauer, Fabian and Gupta, Rahul and Go, Dongwook and Mokrousov, Yuriy and Jakob, Gerhard and Kl\"aui, Mathias},
  journal = {Phys. Rev. B},
  volume = {107},
  issue = {13},
  pages = {134423},
  numpages = {8},
  year = {2023},
  month = {Apr},
  publisher = {American Physical Society},
  doi = {10.1103/PhysRevB.107.134423},
  url = {https://link.aps.org/doi/10.1103/PhysRevB.107.134423}
}
\end{document}